\documentclass[12pt,leqno]{article}
\usepackage{graphicx,psfrag,amssymb,amsmath,amsthm}
 \usepackage[dvips]{color}

\newtheorem{theorem}{Theorem}
\newtheorem{lemma}[theorem]{Lemma}

\def\beq{\begin{equation}}
\def\eeq{\end{equation}}

\def\cB{\mathcal{B}}
\def\cC{\mathcal{C}}

\def\cF{\mathcal{F}}

\def\cK{\mathcal{K}}

\def\cM{\mathcal{M}}

\def\cO{\mathcal{O}}

\def\cR{\mathcal{R}}
\def\cS{\mathcal{S}}

\def\IR{{\mathbb R}}

\def\pQ{\partial Q}

\numberwithin{equation}{section}

\begin{document}

\title{Billiards with polynomial mixing rates}

\author{ N.\ Chernov$^1$ and H.-K.\ Zhang$^1$}

\date{\today}

\maketitle

\footnotetext[1]{Department of Mathematics, University of Alabama
at Birmingham;\\ Email:$\ $ chernov@math.uab.edu;
zhang@math.uab.edu}

\begin{abstract}
While many dynamical systems of mechanical origin, in particular
billiards, are strongly chaotic -- enjoy exponential mixing, the
rates of mixing in many other models are slow (algebraic, or
polynomial). The dynamics in the latter are intermittent between
regular and chaotic, which makes them particularly interesting in
physical studies. However, mathematical methods for the analysis
of systems with slow mixing rates were developed just recently and
are still difficult to apply to realistic models. Here we reduce
those methods to a practical scheme that allows us to obtain a
nearly optimal bound on mixing rates. We demonstrate how the
method works by applying it to several classes of chaotic
billiards with slow mixing as well as discuss a few examples where
the method, in its present form, fails.
\end{abstract}

\section{Introduction}
\label{secI}

A billiard is a mechanical system in which a point particle moves
in a compact container $Q$ and bounces off its boundary $\pQ$. It
preserves a uniform measure on its phase space, and the
corresponding collision map (generated by the collisions of the
particle with $\pQ$, see below) preserves a natural (and often
unique) absolutely continuous measure on the collision space. The
dynamical behavior of a billiard is determined by the shape of the
boundary $\pQ$, and it may vary greatly from completely regular
(integrable) to strongly chaotic.

In this paper we only consider planar billiards, where $Q \subset
\IR^2$. The dynamics in simple containers (circles, ellipses,
rectangles) are completely integrable. The first class of chaotic
billiards was introduced by Ya.~Sinai in 1970 \cite{Si70}; he
proved that if $\pQ$ is convex inward and there are no cusps on
the boundary, then the dynamics is hyperbolic (has no zero
Lyapunov exponents), ergodic, mixing and K-mixing. He called such
systems \emph{dispersing billiards}, now they are often called
\emph{Sinai billiards}. Gallavotti and Ornstein \cite{GO} proved
in 1976 that Sinai billiards are Bernoulli systems.

Many other classes of planar chaotic billiards have been found by
Bunimovich \cite{Bu74,Bu79} in the 1970s, and by Wojtkowski
\cite{W86}, Markarian \cite{M88}, Donnay \cite{D}, and again
Bunimovich \cite{Bu91} in the 1980s and early 1990s. All of them
are proven to be hyperbolic, and some -- ergodic, mixing, and
Bernoulli.

Multidimensional chaotic billiards are known as well, they include
two classical models of mathematical physics -- periodic Lorentz
gases and hard ball gases, for which strong ergodic properties
have been established in a series of fundamental works within the
last 20 years, we refer the reader to a recent collection of
surveys \cite{HB}.

However, ergodic and mixing systems (even Bernoulli systems) may
have quite different statistical properties depending on the rate
mixing (the rate of the decay of correlations), whose precise
definition is given below. On the one hand, the strongest chaotic
systems -- Anosov diffeomorphisms and expanding interval maps --
have exponential mixing rates, see a survey \cite{CY}. This fact
implies the central limit theorem, the convergence to a Brownian
motion in a proper space-time limit, and many other useful
approximations by stochastic processes that play crucial roles in
statistical mechanics.

On the other hand, many hyperbolic, ergodic and Bernoulli systems
have slow (polynomial) mixing rates, which cause weak statistical
properties. Even the central limit theorem may fail, see again a
survey \cite{CY}. Such systems are, in a sense,
\emph{intermittent}, they exemplify a delicate transition from
regular behavior to chaos. For this reason they have attracted
considerable interest in mathematical physics community during the
past 20 years.

The rates of mixing in chaotic billiards is rather difficult to
establish, though, because the dynamics has singularities, which
aggravate the analysis and make standard approaches (based on
Markov partitions and transfer operators) inapplicable. For planar
Sinai billiards, L.-S.~Young \cite{Y98} developed in 1998 a novel
method (now known as \emph{Young's tower construction}) to prove
exponential (fast) mixing rates. Young applied it to Sinai
billiards under two restrictions -- no corner points on the
boundary and finite horizon. Her method was later extended to all
planar Sinai billiards \cite{C99} and to more general
billiard-like Hamiltonian systems \cite{C01,BT}.

There are no rigorous results on the decay of correlations for
multidimensional chaotic billiards yet, since even the best
methods cannot handle a recently discovered phenomenon
characteristic for billiards in dimension 3 and higher -- the
blow-up of the curvature of singularity manifolds \cite{BCST}.

For planar billiards with slow (nonexponential) mixing rates, very
little is known. They turned out to be even harder to treat than
Sinai billiards. Most published accounts are based on numerical
experiments and heuristic analysis, which suggest that Sinai
billiards with cusps on the boundary \cite{Ma} and Bunimovich
billiards \cite{VCG} have polynomial mixing rates. Slow mixing in
these models appears to be induced by tiny places in the phase
space, where the motion is nearly regular, and where the moving
particle can be caught and trapped for arbitrarily long times.

A general approach to the studies of abstract hyperbolic systems
with slow mixing rates was developed by Young in 1998 \cite{Y98},
but its application to billiards required substantial extra
effort, and it took five more years. Only in 2004 R.~Markarian
\cite{M04} used Young's method to establish polynomial mixing
rates for one class of chaotic billiards -- Bunimovich stadia. He
showed that the correlations for the corresponding collision map
decayed as $\cO \bigl( n^{-1} \ln^2n \bigr)$.

In this paper we generalize and simplify the method due to Young
and Markarian essentially reducing it to one key estimate that
needs to be verified for each chaotic billiard table. This
estimate, see (\ref{SS}) in Section~\ref{SecPS}, has explicit
geometric meaning and can be checked by rather straightforward
(but sometimes lengthy) computations, almost in an algorithm-like
manner. We also remark that the method is not restricted to
billiards, it is designed for general nonuniformly hyperbolic
systems with Sinai-Ruelle-Bowen (SRB) measures.

Then we apply the above method to several classes of billiards.
These include semi-dispersing billiards in rectangles with
internal scatterers, Bunimovich flower-like regions, and
``skewed'' stadia (``drive-belt'' tables). In each case we prove
that the correlations decay as $\cO \bigl( n^{-a} \ln^{1+a}n
\bigr)$ for a certain $a>0$, which depends on the character of
``traps'' in the dynamics, and is different for different tables.
We also show that for some billiard tables our key estimate fails,
hence the correlation analysis requires further work.

\section{Statement of results}
\label{secSR}

Here we state our estimates on the decay of correlations for
several classes of billiard systems. The general method for
proving these estimates is described in the next two sections.

First we recall standard definitions \cite{BSC90,BSC91,C97,C99}. A
billiard is a dynamical system where a point moves freely at unit
speed in a domain $Q$ ({\em the table}) and reflects off its
boundary $\pQ$ ({\em the wall}) by the rule ``the angle of
incidence equals the angle of reflection''. We assume that
$Q\subset\IR^2$ and $\pQ$ is a finite union of $C^3$ curves
(arcs). The phase space of this system is a three dimensional
manifold $Q\times S^1$. The dynamics preserves a uniform measure
on $Q\times S^1$.

\begin{figure}[h]
\centering
\centering
\psfrag{r}{$r$}
\psfrag{p}{$\varphi$}
\psfrag{1}{$\frac{\pi}{2}$}
\psfrag{2}{$-\frac{\pi}{2}$}
\psfrag{n}{$n$}
\psfrag{a}{$(a)$}
\psfrag{b}{$(b)$}
\includegraphics[height=1.5in]{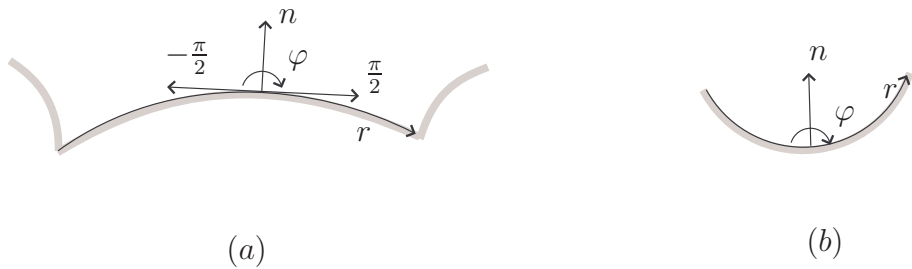}
\renewcommand{\figurename}{Fig.}
\caption{Orientation of $r$ and $\varphi$}\label{rp}
\end{figure}

Let $\cM=\pQ\times [-\pi/2,\pi/2]$ be the standard cross-section
of the billiard dynamics, we call $\cM$ the {\em collision space}.
Canonical coordinates on $\cM$ are $r$ and $\varphi$, where $r$ is
the arc length parameter on $\pQ$ and $\varphi\in [-\pi/2,\pi/2]$
is the angle of reflection, see Fig.~\ref{rp}.

 The first return map
$\cF:\cM\to \cM$ is called the {\em collision map} or the {\em
billiard map}, it preserves smooth measure $d\mu=\cos\varphi\,
dr\, d\varphi$ on $\cM$.

Let $f,g\in L^2_{\mu}(\cM)$ be two functions. {\em Correlations}
are defined by
\beq
   \cC_n(f,g,\cF,\mu) = \int_{\cM} (f\circ \cF^n)\, g\, d\mu -
    \int_{\cM} f\, d\mu    \int_{\cM} g\, d\mu
       \label{Cn}
\eeq
It is well known that $\cF:\cM\to\cM$ is {\em mixing} if and only
if
\beq \label{Cto0}
   \lim_{n\to\infty}
   \cC_n(f,g,\cF,\mu) = 0
   \qquad \forall  f,g\in L^2_{\mu}(\cM)
\eeq
The rate of mixing of $\cF$ is characterized by the speed of
convergence in (\ref{Cto0}) for smooth enough functions $f$ and
$g$. We will always assume that $f$ and $g$ are H\"older
continuous or piecewise H\"older continuous with singularities
that coincide with those of the map $\cF^k$ for some $k$. For
example, the free path between successive reflections is one such
function.

We say that correlations decay {\em exponentially} if
$$
    |\cC_n(f,g,\cF,\mu)|<\,{\rm const}\cdot e^{-cn}
$$
for some $c>0$ and {\em polynomially} if
$$
   |\cC_n(f,g,\cF,\mu)|<\,{\rm const}\cdot n^{-a}
$$
for some $a>0$. Here the constant factor depends on $f$ and $g$.

Next we state our results.

\begin{figure}[h]
\centering
\begin{minipage}[c]{0.3\linewidth}
\psfrag{x}[][]{\small$x$} \psfrag{y}[][]{\small$y$}
\psfrag{a}[][]{\scriptsize$\cB_1$} \psfrag{b}[][]{\scriptsize
$\cB_2$} \psfrag{Q}[][]{\scriptsize $Q$} \psfrag{1}{$(a)$}
\includegraphics[height=1.5in, width=1.3in]{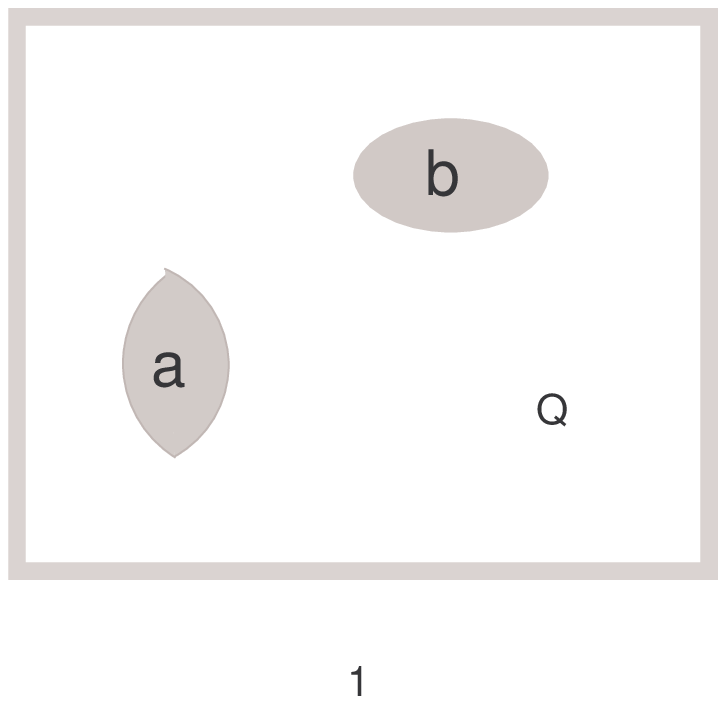}
\hfill
\renewcommand{\figurename}{Fig.}
\end{minipage}%
\begin{minipage}[c]{0.3\linewidth}

\psfrag{Q}[][]{\scriptsize$Q$}
 \psfrag{2}{$(b)$}
\includegraphics[height=1.5in]{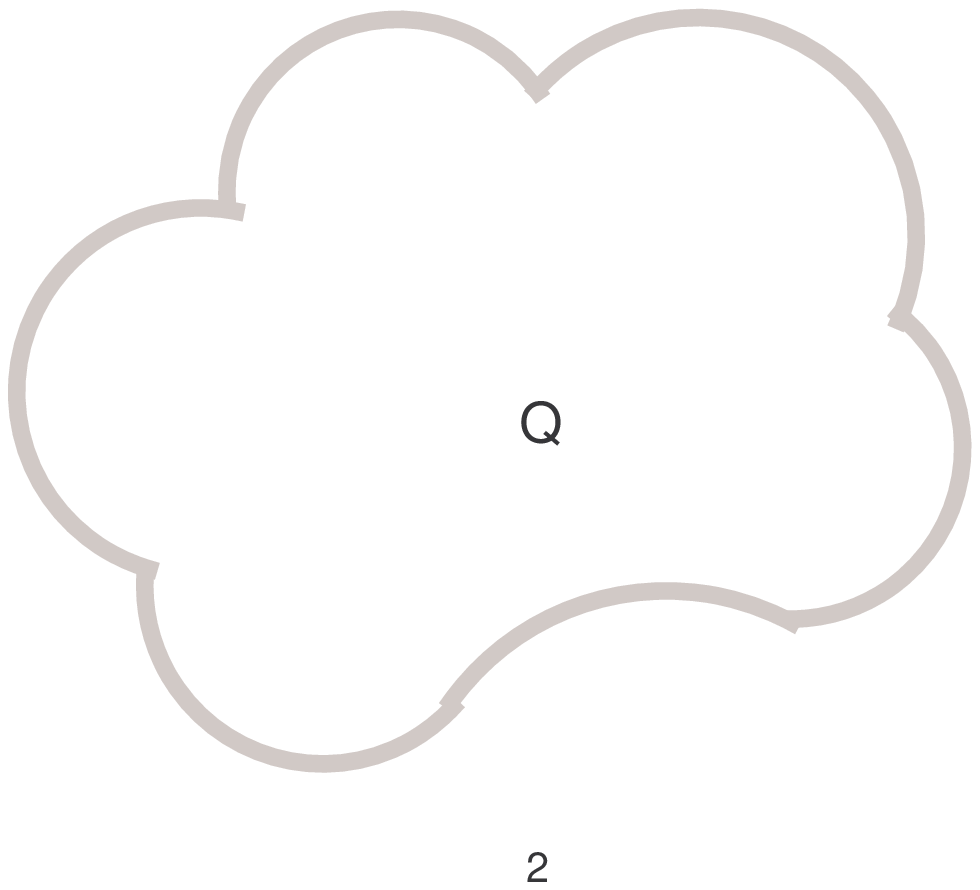}

\renewcommand{\figurename}{Fig.}
\end{minipage}
\begin{minipage}[c]{0.3\linewidth}
\centering
\psfrag{A}[lb][lb]{\small$A$}
\psfrag{B}[lt][lt]{\small$B$}
\psfrag{C}[lb][lb]{\small$C$}
\psfrag{D}[lb][lb]{\small $D$}
\psfrag{3}{$(c)$}
\includegraphics[height=1.5in]{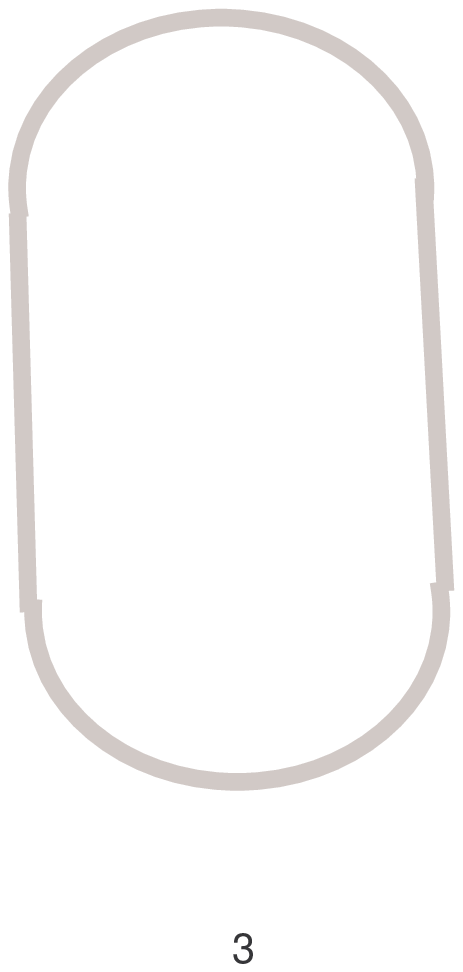}
\renewcommand{\figurename}{Fig.}
\end{minipage}
\renewcommand{\figurename}{Fig.}
\caption{Slow mixing billiard tables}\label{semi}

\end{figure}

Let $\cR \subset \IR^2$ be a rectangle and
$\cB_1,\ldots,\cB_r\subset\,{\rm int}\, \cR$ open strictly convex
subdomains with smooth or piecewise smooth boundaries whose
curvature is bounded away from zero and such that $\overline{\cB}_i
\cap \overline{\cB}_j =\emptyset$ for $i\neq j$, see Fig.~\ref{semi}
(a). A billiard in $Q = \cR \setminus \cup_i \cB_i$ is said to be
{\em semi-dispersing} since its boundary is partially dispersing
(convex) and partially neutral (flat); the flat part is
$\partial\cR$.

\begin{theorem}
For the above semi-dispersing billiard tables, the correlations
(\ref{Cn}) for the billiard map $\cF:\cM\to\cM$ and piecewise
H\"older continuous functions $f,g$ on $\cM$ decay as
$|\cC_n(f,g,\cF,\mu)|\leq\,{\rm const}\cdot (\ln n)^2/n$.
\end{theorem}

Next, let $Q\subset\IR^2$ be a domain with a piecewise smooth
boundary
$$
      \pQ = \Gamma = \Gamma_1\cup\cdots\cup\Gamma_r
$$
such that each smooth component $\Gamma_i\subset\pQ$ is either
convex inward (dispersing) or convex outward (focusing). Assume
that every focusing component $\Gamma_i$ is an arc of a circle
such that there are no points of $\pQ$ on that circle or inside
it, other than the arc $\Gamma_i$ itself, see Fig.~\ref{semi} (b).
Billiards of this type were introduced by Bunimovich \cite{Bu79}
who established their hyperbolicity and ergodicity.

We make two additional assumptions: (i) if two dispersing
components $\Gamma_i,\Gamma_j$ have a common endpoint, then they
are transversal to each other at that point (no cusps!); (ii)
every focusing arc $\Gamma_i\subset \pQ$ is shorter than half a
circle. We also assume that the billiard table is generic to avoid
certain technical complications, see Section~\ref{SecTm2}.

\begin{theorem}
For the above Bunimovich-type billiard tables, the correlations
(\ref{Cn}) for the billiard map $\cF:\cM\to\cM$ and piecewise
H\"older continuous functions $f,g$ on $\cM$ decay as
$|\cC_n(f,g,\cF,\mu)|\leq\,{\rm const}\cdot (\ln n)^3/n^2$.
\end{theorem}

In fact, the boundary $\pQ$ may have flat components as long as
there is an upper bound on the number of consecutive reflections
off the flat components. For example, this is the case when $\pQ$
has a single flat component or exactly two nonparallel flat
components. Then the above theorem remains valid. However, if the
billiard particle can experience arbitrarily many consecutive
collisions with flat boundaries, then, generally speaking, even
the ergodicity cannot be guaranteed, let alone mixing or decay of
correlations.

Lastly, we consider a special case -- a stadium. It is a convex
domain $Q$ bounded by two circular arcs and two straight lines
tangent to the arcs at their common endpoints, see Fig.~\ref{semi}
(c). We distinguish between a ``straight'' stadium, whose flat
sides are parallel, see Fig.~\ref{semi} (c), and a ``skewed''
stadium, whose flat sides are not parallel, we call them {\em
drive-belt tables}, see Fig.~\ref{unstadium2} (a). Notice that
drive-belt tables contain an arc longer than a half circle, so
they do not satisfy the assumption (ii) of the previous theorem
(or the assumptions made by Markarian in \cite{M04}). Again
billiards of this type were introduced by Bunimovich \cite{Bu79}
who established their hyperbolicity and ergodicity.

\begin{theorem}
For both types of stadia, the correlations (\ref{Cn}) for the
billiard map $\cF:\cM\to\cM$ and piecewise H\"older continuous
functions $f,g$ on $\cM$ decay as $|\cC_n(f,g,\cF,\mu)|\leq\,{\rm
const}\cdot (\ln n)^2/n$.
\end{theorem}

The same bound on correlations for straight stadia is already
obtained by Markarian \cite{M04}, but we include it here for the
sake of completeness.

\section{Correlation analysis}
\label{secGMCA}

Here we present a general method for estimating correlations in
nonuniformly hyperbolic dynamical systems. It is based on Young's
recent results \cite{Y98,Y99} and their extensions by Markarian
\cite{M04} and one of us \cite{C99}.

Let $\cF:\cM\to\cM$ be a nonuniformly hyperbolic map acting on a
Riemannian manifold $\cM$ with or without boundary. We assume that
$\cF$ preserves an ergodic Sinai-Ruelle-Bowen (SRB) measure $\mu$,
see \cite{Y98,C99} for definitions and basic facts.

Young \cite{Y98,Y99} found sufficient conditions under which
correlations for the map $\cF$ decay exponentially. We only sketch
Young's method here, skipping many technical details, in order to
focus on the main condition, see (\ref{Yexp}) below.

The key element of Young's construction is a ``horseshoe''
$\Delta_0$ (a set with a hyperbolic structure, often called a
rectangle). By iterating points $x\in\Delta_0$ under the map $\cF$
until they make proper returns to $\Delta_0$, see definitions in
\cite{Y98}, Young constructs a tower, $\Delta$, in which the
rectangle $\Delta_0$ constitutes the first level (the base). The
induced map $\cF_{\Delta}$ on the tower $\Delta$ moves every point
one level up until it hits the ceiling upon which it falls onto
the base $\Delta_0$. Now, for $x\in\Delta$, let $R(x;\cF,\Delta_0)
= \min\{ k\geq 1:\ \cF_{\Delta}^k(x) \in \Delta_0\}$ denote the
``return time'' of the point $x$ to the base $\Delta_0$. The tower
has infinitely many levels, hence $R$ is unbounded. Young proves
that if the probability of long returns is exponentially small,
then correlations decay exponentially fast. Precisely, if
\beq
    \mu\bigl(x\in\Delta:\ R(x;\cF,\Delta_0)>n\bigr)\leq
    \,{\rm const}\cdot\theta^n\quad\quad \forall n\geq 1
       \label{Yexp}
\eeq
where $\theta<1$ is a constant, then $|\cC_n(f,g,\cF,\mu)|<\,{\rm
const}\cdot e^{-cn}$ for some $c>0$.

Therefore, if one wants to derive an exponential bound on
correlations for a particular map, one needs to construct a
``horseshoe'' $\Delta_0$ and verify the tail bound (\ref{Yexp}).
The latter involves all the iterates of the map $\cF$. To simplify
this verification one of us (NC) proposed \cite{C99} sufficient
conditions, which involve just one iterate of the map $\cF$ (or
one properly selected power $\cF^m$, see the next section), that
imply the tail bound (\ref{Yexp}). Those conditions do not require
a construction of a horseshoe $\Delta_0$. We present those
conditions fully in the next section.

Young later extended the results of \cite{Y98} to cover system
with slow mixing rates. She proved \cite{Y99} that if
\beq
    \mu\bigl(x\in\Delta:\ R(x;\cF,\Delta_0)>n\bigr)\leq
    \,{\rm const}\cdot n^{-a}\quad\quad \forall n\geq 1
       \label{Ypol1}
\eeq
where $a>0$ is a constant, then correlations decay polynomially:
\beq
       |\cC_n(f,g,\cF,\mu)|<\,{\rm const}\cdot n^{-a}
         \label{Ypol2}
\eeq

Again, a direct verification of the polynomial tail bound
(\ref{Ypol1}) in specific systems involves all the iterations of
the map $\cF$, and thus might be difficult. There are no known
simplifications, like the one mentioned above, that would reduce
this problem to just one iterate of the map $\cF$ either. However,
there is a roundabout way proposed by Markarian \cite{M04} which
simplifies the analysis, even though it leads to a slightly less
than optimal bound on correlations. We describe Markarian's
approach next.

Suppose one can localize places on the manifold $\cM$ where the
dynamics fails to be strongly hyperbolic, and find a subset
$M\subset\cM$ on which $\cF$ has a strong hyperbolic behavior.
This means, precisely, that the first return map $F:M\to M$
satisfies the conditions of Young's paper \cite{Y98}; in
particular, there exists a horseshoe $\Delta_0\subset M$ for which
an exponential tail bound (\ref{Yexp}) holds for the return times
under the map $F$. Now, of course, the verification of the bound
(\ref{Yexp}) for the map $F$ can be done via the simplified
conditions of \cite{C99}.

Next, consider the return times to $M$ under the original map
$\cF$, i.e.\
\beq
     R(x;\cF,M)=\min\{r\geq 1:\ \cF^r(x)\in M\}
       \label{R0}
\eeq
for $x\in \cM$. Suppose they satisfy a polynomial tail bound
\beq
    \mu(x\in \cM:\ R(x;\cF,M)>n)\leq
    \, {\rm const}\cdot n^{-a}\quad\quad \forall n\geq 1
       \label{Mpol1}
\eeq
where $a>0$ is a constant. Equivalently, we may suppose that
\beq
    \mu(x\in M:\ R(x;\cF,M)>n)\leq
    \, {\rm const}\cdot n^{-a-1}\quad\quad \forall n\geq 1
       \label{Mpol2}
\eeq
The equivalence of (\ref{Mpol1}) and (\ref{Mpol2}) is a simple
fact, we omit its proof. Tail bounds for the return times
$R(x;\cF,M)$ to $M$ are much easier to obtain, in many systems,
than those for the return times $R(x;\cF,\Delta_0)$ to $\Delta_0$.
The following theorem was essentially proved in \cite{M04}, but we
provide a proof here for completeness:

\begin{theorem} \label{tmpolycorr}
Let $\cF:\cM\to \cM$ be a nonuniformly hyperbolic map. Suppose
$M\subset \cM$ is a subset such that the first return map $F:M\to
M$ satisfies the tail bound (\ref{Yexp}) for the return times
$R(x;F,\Delta_0)$ to a rectangle $\Delta_0\subset M$. If the
return times $R(x;\cF,M)$ satisfy the polynomial bound
(\ref{Mpol1}) or, equivalently, (\ref{Mpol2}), then
\beq
       |\cC_n(f,g,\cF,\mu)|<\,{\rm const}\cdot
       (\ln n)^{a+1} n^{-a}
         \label{Mpol3}
\eeq
\end{theorem}

\noindent{\em Proof}. For every $n\geq 1$ and $x\in\cM$ denote
$$
  r(x;n,M)=\#\{1\leq i\leq n:\ \cF^i(x)\in M\}
$$
and
\begin{align*}
    A_n &=\{x\in \cM\colon\ R(x;\cF,\Delta_0)>n\},\\
    B_{n,b} &=\{x\in \cM\colon\ r(x;n,M) > b\ln n\},
\end{align*}
where $b>0$ is a constant to be chosen shortly. By (\ref{Yexp}),
$$
    \mu(A_n \cap B_{n,b})\leq\,{\rm const}\cdot n\,\theta^{b\ln n}.
$$
Choosing $b$ large enough makes this bound less than const$\cdot
n^{-a}$.

To bound $ \mu(A_n \setminus B_{n,b})$ we note that points $x\in
A_n \setminus B_{n,b}$ return to $M$ at most $b\ln n$ times during
the first $n$ iterates of $\cF$. In other words, there are $\leq
b\ln n$ time intervals between successive returns to $M$, and
hence the longest such interval, we call it $I$, has length $\geq
n/(b\ln n)$. Applying the bound (\ref{Mpol2}) to the interval $I$
gives
$$
     \mu(A_n \setminus B_{n,b})
     \leq \,{\rm const}\cdot n\,(\ln n)^{a+1}/n^{a+1}
$$
(the extra factor of $n$ must be included because the interval $I$
may appear anywhere within the longer interval $[1,n]$, and the
measure $\mu$ is invariant). In terms of Young's tower $\Delta$,
we obtain
\beq
    \mu(x\in\Delta:\ R(x;\cF,\Delta_0)>n)\leq
    \,{\rm const}\cdot (\ln n)^{a+1}\, n^{-a}\quad\quad \forall n\geq 1
       \label{Ypol3}
\eeq
This tail bound differs from Young's (\ref{Ypol1}) by the extra
factor $(\ln n)^{a+1}$. However, as it was explained by Markarian
\cite{M04}, the same argument that Young used to derive
(\ref{Ypol2}) from (\ref{Ypol1}) now gives us (\ref{Mpol3}) based
on (\ref{Ypol3}). This completes the proof of the theorem. \qed

\section{Conditions for exponential mixing}
\label{secSCED}

Here we list sufficient conditions on a 2D nonuniformly hyperbolic
map $F\colon M \to M$ with a mixing SRB measure $\mu$ under which
its correlations decay exponentially. These are a 2D version of
more general conditions stated in \cite{C99}. We also provide
comments that will help us apply these conditions to chaotic
billiards.

We assume that $M$ is an open domain in a two-dimensional
$C^{\infty}$ compact Riemannian manifold $\cM$ with or without
boundary. For any smooth curve $W\subset\cM$ we denote by $\nu_W$
the Lebesgue measure on $W$ (induced by the Euclidean metric). For
brevity, $|W| = \nu_W(W)$ will denote the length of $W$.

\bigskip\noindent {\bf 4.1 Smoothness.}
The map $F$ is a $C^2$ diffeomorphism of $M \setminus {\cS}$ onto
$F(M \setminus {\cS})$, where $\cS$ is a closed set of zero
Lebesgue measure. Usually, $\cS$ is the set of points at which $F$
either is not defined or is singular (discontinuous or not
differentiable).

\medskip \emph{Remark}.
The collision map $\cF\colon \cM \to \cM$ for a billiard table
with a piecewise smooth $C^r$ boundary is piecewise $C^{r-1}$
smooth \cite{CM}. The singularities of $\cF$ make a closed set
$\cS$ that is a finite or countable union of smooth compact curves
in the collision space $\cM$. On those curves, one-sided
derivatives of $\cF$ are often infinite. When we construct a first
return map $F\colon M\to M$ on a subdomain $M \subset \cM$, it
will be always clear that $F$ has similar properties.

We denote by $\cS_m = \cS \cup \cF^{-1}(\cS) \cup \cdots \cup
\cF^{-m+1}(\cS)$ the singularity set for the map $\cF^m$.
Similarly, $\cS_{-m}$ denotes the singularity set for the map
$\cF^{-m}$.

\bigskip\noindent {\bf 4.2 Hyperbolicity}.
There exist two families of cones $C^u_x$ (unstable) and $C^s_x$
(stable) in the tangent spaces ${\cal T}_xM$, $x\in \bar{M}$, such
that $DF(C^u_x)\subset C^u_{Fx}$ and $DF(C^s_x)\supset C^s_{Fx}$
whenever $DF$ exists, and
$$
     |DF(v)|\geq \Lambda|v|\quad\forall v\in C^u_x
      \quad\text{and}\quad
     |DF^{-1}(v)|\geq \Lambda|v|\quad\forall v\in C^s_x
$$
with some constant $\Lambda>1$. Here $|\cdot|$ is some norm
equivalent (though not necessarily uniformly equivalent) to the
Euclidean norm. These families of cones are continuous on
$\bar{M}$ and the angle between $C^u_x$ and $C^s_x$ is bounded
away from zero.

For any $F$-invariant measure $\mu'$, almost every point $x\in M$
has one positive and one negative Lyapunov exponent. Also, almost
every point $x$ has one-dimensional local unstable and stable
manifolds, which we denote by $W^u(x)$ and $W^s(x)$, respectively.

\medskip \emph{Remark}. The existence of Lyapunov exponents usually
follows from the Oseledec theorem, and the existence of stable and
unstable manifolds -- from the Katok-Strelcyn theorem, see
\cite{KS}. Both theorems require certain mild technical
conditions, which have been verified for virtually all planar
billiards in \cite{KS}. The hyperbolicity has been proven for all
the classes of billiards studied in this paper. It is also known
that the tangent vectors to the curves in $\cS$ lie in stable
cones, and those of the curves in $\cS_{-1}$ -- in unstable cones.

\bigskip\noindent {\bf 4.3 SRB measure.}
The map $F$ preserves an ergodic measure $\mu$ whose conditional
distributions on unstable manifolds are absolutely continuous.
Such a measure is called a Sinai-Ruelle-Bowen (SRB) measure. We
also assume that $\mu$ is mixing.

\medskip\emph{Remark}.
For the collision map of a chaotic billiard system, the natural
smooth invariant measure $\mu$ is an SRB measure \cite{Y98}. Its
ergodicity and mixing have been proven for all classes of
billiards studied in this paper, see references in
Section~\ref{secSR}.

\bigskip\noindent {\bf 4.4 Distortion bounds.}
Let $\Lambda(x)$ denote the factor of expansion on the unstable
manifold $W^u(x)$ at the point $x$. If $x,y$ belong to one
unstable manifold $W^u$ such that $F^n$ is defined and smooth on
$W^u$, then
\beq
     \log\prod_{i=0}^{n-1}\frac{\Lambda(F^ix)}{\Lambda(F^iy)}
       \leq \psi\bigl({\rm dist}(F^nx,F^ny)\bigr)
           \label{distor1}
\eeq
where $\psi(\cdot )$ is some function, independent of $W$, such
that $\psi(s)\to 0$ as $s\to 0$.

\medskip\emph{Remark}.
Since the derivatives of the collision map $\cF$ turn infinite on
some singularity curves $\cS$, the expansion of unstable manifolds
terminating on those curves is highly nonuniform. To enforce the
required distortion bound, one follows a standard procedure
introduced in \cite{BSC91}, see also \cite{Y98,C99}. Let $\cM_+$
denote the part of the collision space corresponding to dispersing
walls $\Gamma_i \subset \partial Q$, i.e.\
$$
   \cM_+ = \{(r, \varphi)\colon\, r\in\Gamma_i,
   \ \Gamma_i\text{ is dispersing}\}.
$$
One divides $\cM_+$ into countably many sections (called
\emph{homogeneity strips}) defined by
$$
    H_k=\{(r,\varphi)\in\cM_+\colon \pi/2-k^{-2}<\varphi <\pi/2-(k+1)^{-2}\}
$$
and
$$
    H_{-k}=\{(r,\varphi)\in\cM_+\colon -\pi/2+(k+1)^{-2}<\varphi < -\pi/2+k^{-2}\}
$$
for all $k\geq k_0$ and
\beq \label{bbH0}
    H_0=\{(r,\varphi)\in\cM_+\colon -\pi/2+k_0^{-2}<\varphi <
    \pi/2-k_0^{-2}\},
\eeq
here $k_0 \geq 1$ is a fixed (and usually large) constant. Then, a
stable (unstable) manifold $W$ is said to be \emph{homogeneous} if
its image $\cF^n(W)$ lies either in one homogeneity strip of
$\cM_+$ or in $\cM \setminus \cM_+$ for every $n\geq 0$ (resp.,
$n\leq 0$). It is shown in \cite{BSC91} that a.e.\ point $x\in
\cM$ has homogeneous stable and unstable manifolds passing through
it. The distortion bounds for homogeneous unstable manifolds were
proved in \cite{BSC91,C99}. It is also easy to check that the
distortion bounds for the collision map $\cF$ imply similar
bounds, with the same function $\psi(s)$, for the induced map
$F\colon M\to M$ on any subdomain $M \subset \cM$.

>From now on, we will only deal with homogeneous manifolds without
even mentioning this explicitly. This can be guaranteed by
redefining the collision space, as it is done in \cite{C99}: we
remove from $\cM_+$ the boundaries $\partial H_k$, thus making
$\cM_+$ a countable disjoint union of the open homogeneity strips
$H_k$'s. (There is no need to subdivide or otherwise redefine $\cM
\setminus \cM_+$, see \cite{C99}) Accordingly, the images
(preimages) of $\partial H_k$ need to be added to the set
$\cS_{-1}$ (resp. $\cS$). We call them \emph{new singularities},
and refer to the original sets $\cS$ and $\cS_{-1}$ as \emph{old
singularities}. Now the stable and unstable manifolds for the map
$\cF$ on the (thus redefined) collision space $\cM$ will be always
homogeneous and the distortion bounds will always hold.
\medskip

\bigskip\noindent {\bf 4.5 Bounded curvature.}
The curvature of unstable manifolds is uniformly bounded by a
constant $B\geq 0$.

\bigskip\noindent {\bf 4.6 Absolute continuity.} If $W_1,W_2$ are two small
unstable manifolds close to each other, then the holonomy map $h:\
W_1\to W_2$ (defined by sliding along stable manifolds) is
absolutely continuous with respect to the Lebesgue measures
$\nu_{W_1}$ and $\nu_{W_2}$, and its Jacobian is bounded, i.e.
\beq
 1/C'\leq\frac{\nu_{W_2}(h(W_1'))}{\nu_{W_1}(W_1')}\leq C'
     \label{ac}
\eeq
with some $C'>0$, where $W_1' \subset W_1$ is the set of points
where $h$ is defined.

\medskip\emph{Remark}.
The properties 4.5 and 4.6 have been verified for planar chaotic
billiards considered in this paper \cite{BSC91,C99} and even for
more general billiard-like Hamiltonian systems \cite{C01}.
\medskip

Before we state the last (and most important) condition, we need
to introduce some notation. Let $\delta_0>0$. We call an unstable
manifold $W$ a $\delta_0$-LUM (local unstable manifold) if $|W|
\leq \delta_0$. Let $V\subset W$ be an open subset, i.e.\ a finite
or countable union of open sumbintervals of $W$. For $x\in V$
denote by $V(x)$ the subinterval of $V$ containing the point $x$.
Let $n\geq 0$. We call an open subset $V\subset W$ a
$(\delta_0,n)$-subset if the map $F^n$ is defined and smooth on
$V$ and $|F^n V(x)| \leq \delta_0$ for every $x\in V$. Note that
$F^n V$ is then a union of $\delta_0$-LUM's. Define a function
$r_{V,n}$ on $V$ by
\beq
    r_{V,n}(x)=\,{\rm dist}(F^nx,\partial F^nV(x)),
      \label{rVn}
\eeq
which is the distance from $F^n(x)$ to the nearest endpoint of the
curve $F^nV(x)$. In particular, $r_{W,0}(x)=\,{\rm dist}
(x,\partial W)$.

\bigskip\noindent {\bf 4.7 One-step growth of unstable manifolds.}
There is a small $\delta_0 >0$ and constants $\alpha_0\in (0,1)$
and $\beta_0,D,\kappa,\sigma>0$ with the following property. For
any sufficiently small $\delta>0$ and any $\delta_0$-LUM $W$
denote by $U^1_{\delta} = U^1_{\delta}(W) \subset W$ the
$\delta$-neighborhood of the set $W\cap \cS$ within $W$. Then
there is an open $(\delta_0 ,1)$-subset $V^1_{\delta} =
V^1_{\delta}(W) \subset W\setminus U^1_{\delta}$ such that
$\nu_W\bigl(W\setminus (U^1_{\delta}\cup V^1_{\delta})\bigr)=0$
and $\forall\varepsilon>0$
\beq
  \nu_W\bigl(r_{V^1_{\delta},1}<\varepsilon\bigr)
  \leq  2\alpha_0\varepsilon +
  \varepsilon\beta_0\delta_0^{-1}|W|
    \label{growth1}
\eeq
\beq
  \nu_W\bigl(r_{U^1_{\delta},0} < \varepsilon\bigr)
  \leq  D\delta^{-\kappa}\varepsilon
    \label{growth2}
\eeq
and
\beq
    \nu_W\bigl(U^1_{\delta}\bigr) \leq D\delta^{\sigma}
      \label{growth3}
\eeq

We comment on these conditions after stating the theorem proved in
\cite{C99}:

\begin{theorem}[\cite{C99}]
Let $F$ satisfy the assumptions 4.1--4.7. Then (a) there is a
horseshoe $\Delta_0\subset M$ such that the return times
$R(x;F,\Delta_0)$ satisfy the tail bound (\ref{Yexp}). Thus, (b)
the map $F:M\to M$ enjoys exponential decay of correlations.
\label{tmmain}
\end{theorem}

\medskip\emph{Remark}.
It is shown in \cite{C99}, Proposition 10.1, that if the
conclusions (a) and (b) of this theorem hold for some power $F^m$,
$m \geq 2$, of the map $F$, then they hold for $F$ itself.
Therefore, it will be enough to verify the condition 4.7 for the
map $F^m$ with some $m \geq 2$ (of course, $\cS$ should then be
replaced by $\cS_m$, $r_{V,1}$ by $r_{V,m}$, etc.). This option
will be important when the condition (\ref{growth1}) fails for the
map $F$ but holds for some power $F^m$, see below. In this case we
do not have to worry about (\ref{growth2}) and (\ref{growth3}),
since they will hold for all $F^m$, $m\geq 1$, see the next
section.

\section{Simplified conditions for exponential mixing}

Here we simplify the conditions (\ref{growth1})--(\ref{growth3})
and reduce them to one inequality that needs to be checked for
every particular class of chaotic billiards.

First, the condition (\ref{growth2}) always holds for 2D maps,
where dim$\, W=1$, as it was shown in \cite{C99}, and the reason
is simple: each connected component of $U^1_{\delta}$ has length
$\geq 2\delta$, thus there are no more than $\delta_0/(2\delta)$
of them, hence $ \nu_W \bigl( r_{U^1_{\delta},0} < \varepsilon
\bigr) \leq \delta_0\delta^{-1}\varepsilon$. Thus, (\ref{growth2})
will hold for all our applications.

The condition (\ref{growth3}) holds under very mild assumptions on
the structure of the set $\cS$, which will be always satisfied in
our applications:

\medskip\noindent{\bf Assumption (Structure of the singularity
set)}. Assume that for any unstable curve $W \subset M$ (i.e., a
curve whose tangent vectors lie in unstable cones) the set $W\cap
\cS$ is finite or countable and has at most $K$ accumulation
points on $W$, where $K \geq 1$ is a constant. Let $x_{\infty}$ be
one of them and $\{x_n\}$ denote the monotonic sequence of points
$W\cap \cS$ converging to $x_{\infty}$. We assume that
\beq \label{Cnd}
         {\rm dist}(x_n,x_{\infty}) \leq C/n^d
\eeq
for some constants $C,d>0$, i.e.\ the convergence of $x_n$ to
$x_{\infty}$ is faster than some power function of $n$.

\begin{lemma} \label{Lm1}
In the notation of 4.7, our assumption (\ref{Cnd}) implies
\beq \label{growth3a}
    \nu_W\bigl(U^1_{\delta}\bigr) <
     4K\,C^{\frac{1}{1+d}}\,\delta^{\frac{d}{1+d}},
\eeq
which in turn implies (\ref{growth3}).
\end{lemma}

\medskip\noindent{\em Proof}.
It is obviously enough to consider the case $K=1$. Let $n_{\ast} =
C^{\frac{1}{1+d}}\delta^{-\frac{1}{1+d}}$ and note that
$U^1_{\delta}(W)$ is contained in the union of the interval
$(x_{\infty},x_{n_{\ast}})$ and $n_{\ast}$ intervals of length
$2\delta$ centered on $x_i$, $1\leq i\leq n_{\ast}$, thus
$$
   \nu_W\bigl(U^1_{\delta}(W)\bigr)
   \leq C/n_{\ast}^d + 2\,n_{\ast}\delta
$$
which easily implies (\ref{growth3a}). \qed \medskip

\begin{lemma} \label{Lm2}
Under our assumption (\ref{Cnd}), the bound (\ref{growth3}) holds
for any power $F^m$ (with $D$ and $\sigma$ depending on $m$, of
course).
\end{lemma}

\medskip\noindent{\em Proof}.
We use induction on $m$. Let $W$ be an unstable manifold and
$U^m_{\delta}(W) \subset W$ denote the $\delta$-neighborhood of
the set $W \cap \cS_m$ within $W$. Assume that (\ref{growth3})
holds, i.e.\
\beq \label{nu2m}
    \nu_W\bigl(U^m_{\delta}(W)\bigr)
    \leq D\delta^{\sigma}
\eeq
for some constants $D,\sigma>0$ and any $\delta >0$. Set
$$
       \delta_{\ast} = \delta^{\frac{d}{(1+d)(1+\sigma)}},
$$
then $\nu_W\bigl(U^m_{\delta_{\ast}}(W)\bigr) \leq D
\delta_{\ast}^{\sigma}$ by (\ref{nu2m}). Let $I_1,\ldots, I_N$ be
the connected components of $W \setminus \cS_m$ whose length is $>
\delta_{\ast}$. Note that $N \leq |W| / \delta_{\ast} \leq
\delta_0 / \delta_{\ast}$. For each $i=1,\ldots,N$ the set $J_i =
F^m(I_i)$ is an unstable manifold. By the assumption (\ref{Cnd}),
the set $J_i \cap \cS$ is at most countable and has $\leq K$
accumulation points which satisfy (\ref{Cnd}). Since $F^{-m}$ is
smooth and contracting on each $J_i$, the set $I_i \cap \cS_{m+1}$
also is at most countable and has $\leq K$ accumulation points
which satisfy (\ref{Cnd}). Now by the same argument as in the
proof of Lemma~\ref{Lm1}
$$
    \nu_W\bigl(U_{\delta}^{m+1}(I_i)\bigr)
    \leq  4KC^{\frac{1}{1+d}}\,\delta^{\frac{d}{1+d}},
$$
for each $i$. Observe that
$$
  U_{\delta}^{m+1}(W) \subset
   U_{\delta_{\ast}}^m(W) \bigcup
   \biggl(\bigcup_{i=1}^N U_{\delta}^{m+1}(I_i)\biggr),
$$
thus
\begin{align*}
  \nu_W\bigl(U_{\delta}^{m+1}(W)\bigr) &\leq
    D\delta_{\ast}^{\sigma} + \,{\rm const}\cdot
    \delta_{\ast}^{-1} \delta^{\frac{d}{1+d}} \\
    &\leq \,{\rm const}\cdot
     \delta^{\frac{d\sigma}{(1+d)(1+\sigma)}},
\end{align*}
which proves (\ref{growth3}) for the map $F^{m+1}$. \qed \medskip

\noindent\emph{Remark S}. We need to extend Lemma~\ref{Lm1} to a
slightly more general type of singularities. Suppose $\cS_1$ and
$\cS_2$ are two sets such that each satisfies the above assumption
(\ref{Cnd}), and $\cS = \cS_1 \cup F^{-1} (\cS_2)$. Then the
argument used in the proof of Lemma~\ref{Lm2} shows that $\cS$
satisfies (\ref{growth3}). Furthermore, in this case $\cS_m$
satisfies (\ref{growth3}) due to the same argument.
\medskip

We now turn to the most important condition (\ref{growth1}). Let
$W_i$ denote the connected components of $W \setminus \cS_1$ and
$$
   \Lambda_i = \min_{x\in W_i} \Lambda(x)
$$
the minimal local expansion factor of the map $F$ on $W_i$. We
note that due to the distortion bounds
\beq \label{LLxy}
   \forall x,y\in W_i\quad\quad
   e^{-\psi(\delta_0)}
   \leq \frac{\Lambda(x)}{\Lambda(y)}
   \leq e^{\psi(\delta_0)},
\eeq
and so due to the smallness of $\delta_0$ the expansion factor is
in fact almost constant on each $W_i$.

\begin{lemma} \label{lm1step}
The condition (\ref{growth1}) is equivalent to
\beq
   \alpha_1\colon = \liminf_{\delta_0\to 0}\
   \sup_{W\colon |W|<\delta_0} \sum_i \Lambda_i^{-1} <1,
      \label{step1}
\eeq
where the supremum is taken over unstable manifolds $W$ and
$\Lambda_i$, $i \geq 1$, denote the minimal local expansion
factors of the connected components of $W \setminus \cS$ under
$F$.
\end{lemma}

We call (\ref{step1}) a \emph{one-step expansion estimate} for
$F$.

\medskip\noindent\emph{Proof}. We prove that (\ref{step1}) implies
(\ref{growth1}), the converse implication will be self-evident in
the end (the converse will not be used in practical applications).

Let $W^1_i = W_i \setminus \overline{U^1_{\delta}(W)}$ be the open
subinterval of $W_i$ obtained by removing from $W_i$ the
$\delta$-neighborhood of its endpoints (of course, if $|W_i|<
2\delta$, then $W_i^1 = \emptyset$), and let $W^1 = \cup_i W^1_i$.
It is easy to see that $\forall \varepsilon >0$
\beq
   \nu_W(r_{W^1,1}<\varepsilon)
     \leq\sum_{i}
     2\varepsilon \Lambda_i^{-1}
     =2\alpha_1\varepsilon
\eeq
We note that this estimate is almost exact, since the expansion
factor is almost constant on each $W_i$ due to (\ref{LLxy}).

Next, to obtain an open $(\delta_0,1)$-subset $V^1_{\delta}
\subset W^1$ as required in (\ref{growth1}), we need to subdivide
the intervals $W^1_i \subset W^1$ into subintervals whose
$F$-images are shorter than $\delta_0$. More precisely, let us
divide each curve $F(W_i^1)$ whose length exceeds $\delta_0$ into
$k_i$ equal subintervals of length between $\delta_0/2$ and
$\delta_0$, with $k_i \leq 2|F(W_i^1)| / \delta_0$. If $|F(W_i^1)|
\leq \delta_0$, then we set $k_i =0$ and leave $W_i^1$ unchanged.
Then the union of the preimages of the $\varepsilon$-neighborhoods
of the new partition points on the curves $F(W_i^1)$ has
$\nu_W$-measure bounded above by
$$
  3\varepsilon \sum_i k_i\, \Lambda_i^{-1}
   \leq 6\varepsilon\,\delta_0^{-1}
   \sum_i |F(W_i^1)|\, \Lambda_i^{-1}
   \leq 7\varepsilon \,\delta_0^{-1}\, |W|
$$
where we increased the numerical coefficient from 6 to 7 in order
to take incorporate the factor $e^{\psi(\delta_0)}$ resulting from
the distortion bounds (\ref{LLxy}). Thus we obtain
\beq
  \nu_W\bigl(r_{V^1_{\delta},1}<\varepsilon\bigr)
  \leq  2\alpha_1\varepsilon +
  7\varepsilon \delta_0^{-1} |W|
    \label{growth1a}
\eeq
Lemma is proved. \qed \medskip

The one-step expansion estimate (\ref{step1}) is geometrically
explicit and easy-to-check, but unfortunately, as stated, it fails
too often. Indeed, suppose $S \subset \cS$ is a singularity curve
that divides an unstable manifold $W$ into two components, $W_1$
and $W_2$. For (\ref{step1}) to hold, we need $\Lambda_1^{-1} +
\Lambda_2^{-1} < 1$, which is a fairly stringent requirement (it
means, in particular, that $\Lambda_1 > 2$ or $\Lambda_2 > 2$).
What if the hyperbolicity of the system at hand is not that
strong, i.e.\ the expansion of unstable curves is less than
twofold? Then (\ref{step1}) fails, and so does (\ref{growth1}).

In that case, according to the remark after Theorem~\ref{tmmain},
it will be enough to prove (\ref{growth1}) for any power $F^m$. If
we apply Lemma~\ref{lm1step} to the map $F^m$ we get the
following:

\begin{lemma} \label{lmmstep}
The analogue of the condition (\ref{growth1}) for $F^m$ is
equivalent to
\beq
   \alpha_m\colon = \liminf_{\delta_0\to 0}\
   \sup_{W\colon |W|<\delta_0} \sum_i \Lambda_{i,m}^{-1} <1,
      \label{stepm}
\eeq
where the supremum is taken over unstable manifolds $W$ and
$\Lambda_{m,i}$, $i \geq 1$, denote the minimal local expansion
factors of the connected components of $W \setminus \cS_m$ under
$F^m$.
\end{lemma}

We call (\ref{stepm}) the \emph{$m$-step expansion estimate}.

To summarize our discussion, we state another theorem:

\begin{theorem}
Let $F$ be defined on a 2D manifold $\cM$ and satisfy the basic
requirements 4.1--4.6. Suppose the singularity set $\cS$ has the
structure described by (\ref{Cnd}) or by Remark~S, and let
(\ref{stepm}) hold. Then (a) there is a horseshoe $\Delta_0\subset
M$ such that the return times $R(x;F,\Delta_0)$ satisfy the tail
bound (\ref{Yexp}). Thus, (b) the map $F:M\to M$ enjoys
exponential decay of correlations.
\end{theorem}

In our applications, 4.1--4.6 will always hold and the singularity
set $\cS$ will obviously have the structure described by
(\ref{Cnd}) or by Remark~S. The only condition we will need to
verify is (\ref{stepm}) for some $m\geq 1$. But still, it involves
a higher iteration of $F$, which might be practically
inconvenient, so we will simplify this condition further in the
next section.

\section{A practical scheme}
\label{SecPS}

The verification of (\ref{stepm}) can be done according to a
general scheme that we outline here.

Suppose $\cS_m$ is a finite union of smooth compact curves that
are uniformly transversal to unstable manifolds (in our
applications, the tangent vectors to those curves lie in stable
cones ensuring transversality). For an unstable manifold $W$, let
$K_m(W)$ denote the number of connected components of $W \setminus
\cS_m$. We call
$$
    K_m = \liminf_{\delta_0 \to 0} \sup_{W\colon |W|<\delta_0}
    K_m(W)
$$
the \emph{complexity of the map} $F^m$. Note that $K_m$ does not
depend on the number of the curves in $\cS_m$, which may grow
rapidly with $m$, but rather on the maximal number of those curves
that meet at any one point of $M$.

\begin{figure}[h]
    \centering
    \psfrag{C}{$C^u_x$}
    \psfrag{S}{$\cS_m$}
    \psfrag{x}{$x$}     \includegraphics[height=1.5in]{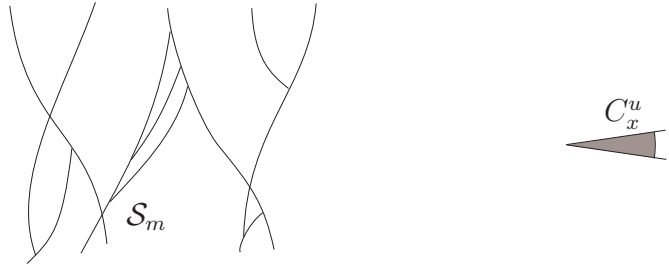}
    \renewcommand{\figurename}{Fig.}
    \caption{Singularity curves and unstable cones.}
    \label{FigSingComp}
\end{figure}

For example, on Fig.~\ref{FigSingComp} a set of 10 curves of
$\cS_m$ is depicted; to ensure transversality, those curves run in
a general vertical direction, while unstable cones (depicted by a
shadowed triangle on the right) are assumed to be nearly
horizontal; the complexity $K_m$ here equals 3, since sufficiently
short unstable manifolds intersect at most 2 curves in $\cS_m$.

\begin{lemma} \label{lmKm}
Suppose that for some $m\geq 1$
\beq \label{complexbound}
       K_m < \Lambda^{m}
\eeq
where $\Lambda > 1$ is the minimum expansion factor of unstable
vectors. Then (\ref{stepm}) holds, thus (\ref{growth1}) holds for
$F^m$.
\end{lemma}

\medskip\noindent\emph{Proof}. By the chain rule, we have
$\Lambda_{m,i} > \Lambda^m$ in (\ref{stepm}), hence the lemma. \qed
\medskip

The condition (\ref{complexbound}) and alike are known as {\em
complexity bounds} in the literature. In some cases, the sequence
$K_m$ has been proven to grow with $m$ at most polynomially, $K_m
= \cO(m^A)$ for some $A>0$, thus (\ref{complexbound}) holds for
all large enough $m$. It is believed that for typical billiard
systems $K_m$ grows slowly (at least more slowly than any
exponential function), but only in few cases exact proofs are
available.

The above arguments are valid if $\cS_m$ is a \emph{finite} union
of smooth compact curves. However, in many billiards systems (and
in all our applications) $\cS_m$ consists of \emph{countably} many
smooth compact curves, which accumulate at some places in the
space $M$. For example, the boundaries of homogeneity strips
introduced in 4.4, converge to two lines $\varphi = \pm \pi/2$,
hence their preimages (included in $\cS$) converge to the
preimages of the lines $\varphi = \pm \pi/2$. Note that since
$\cS_m$ is closed, the accumulation points of singularity curves
belong to $\cS_m$ as well (they may be single points or some
curves of $\cS_m$).

In this case the singularity curves $\cS$ of the map $F$ can be
usually divided into two groups: a finite number of {\em primary}
curves and finitely or countably many sequences of {\em secondary}
curves, each sequence converges to a limit point on a primary
curve or to a primary curve in $\cS$. In each example, the
partition into primary and secondary singularity curves will be
quite natural.

Now let $W$ be a short unstable manifold that crosses no primary
singularity curves but finitely or countably many secondary
singularity curves. Then $W \setminus \cS$ consists of finitely or
countably many components $W_i$, and we denote by $\Lambda_i$, as
before, the minimum local expansion factor of $W_i$ under the map
$F$. In all our applications, the accumulation of secondary
singularity curves is accompanied by very strong expansion, so
that $\Lambda_i$ are very large. The following is our crucial
assumption:

\medskip\noindent{\bf Assumption (on secondary singularities)}.
We have
\beq
   \theta_0\colon = \liminf_{\delta_0\to 0}\
   \sup_{W\colon |W|<\delta_0} \sum_i \Lambda_i^{-1} <1,
      \label{SS}
\eeq
where the supremum is taken over unstable manifolds $W$
intersecting \emph{no primary singularity curves} and $\Lambda_i$,
$i \geq 1$, denote the minimal local expansion factors of the
connected components of $W \setminus \cS$ under the map $F$.

We denote
$$
        \theta_1\colon
        = \max \{ \theta_0, \Lambda^{-1}\} < 1.
$$

For the primary singularity curves, we assume an analogue of the
complexity bound (\ref{complexbound}) as follows. Let $\cS_{\rm
P}$ be the union of the primary singularity curves and $\cS_{{\rm
P},m} = \cS_{\rm P} \cup F^{-1}\cS_{\rm P} \cup \cdots \cup
F^{-m+1} \cS_{\rm P}$. For an unstable manifold $W$, let $K_{{\rm
P},m}(W)$ denote the number of connected components of $W
\setminus \cS_{{\rm P},m}$. i.e.\ the number of components into
which $W$ is divided by {\em primary singularities only} during
the first $m$ iterations of $F$. We call
$$
    K_{{\rm P},m} = \liminf_{\delta_0 \to 0} \sup_{W\colon |W|<\delta_0}
    K_{{\rm P},m}(W)
$$
the \emph{complexity of the primary singularities of} $F^m$. We
now make our last assumption:

\medskip\noindent{\bf Assumption (on primary singularities)}.
For some $m\geq 1$
\beq \label{Primary}
       K_{{\rm P},m} < \theta_1^{-m}
\eeq

\begin{theorem} \label{Tm11}
Assume (\ref{SS}). If (\ref{Primary}) holds for some $m\geq 1$,
then (\ref{stepm}) follows, and thus (\ref{growth3}) is true for
the map $F^m$.
\end{theorem}

\proof Let $W' \subset W$ be a connected component of the set $W
\setminus \cS_{{\rm P},m}$ and $W_j'$, $j \geq 1$, denote all the
connected components of $W' \setminus \cS_m$. Denote by
$\Lambda_j'$ the minimum expansion factor of $W_j'$ under the map
$F^m$.

\begin{lemma} \label{lmW'}
In the above notation
\beq \label{W'}
         \liminf_{\delta_0\to 0}\
   \sup_{W\colon |W|<\delta_0} \sup_{W'\subset W}
   \sum_j [\Lambda_j']^{-1} <
   \theta_1^m.
\eeq
\end{lemma}

\proof During the first $m$ iterations of $F$, the images of $W'$
never cross any primary singularity curves but may be divided by
secondary ones into finitely or countably many pieces.

Now the proof goes by induction on $m$. For $m=1$, the statement
follows from (\ref{SS}) if $W'$ intersects $\cS$, or from the mere
hyperbolicity 4.2 otherwise. Next we assume (\ref{W'}) for some
$m$ and apply this same argument to each connected component of
the set $F^m(W')$ and then use the chain rule to derive (\ref{W'})
for $m+1$.  \qed
\medskip

Lemma~\ref{lmW'} and the assumption (\ref{Primary}) imply
$$
    \alpha_m \leq K_{{\rm P},m}\theta_1^{m} < 1
$$
which proves the theorem. \qed \medskip

In the following sections, we apply our methods to three classes
of chaotic billiards with slow mixing rates.

\section{Proof of Theorem 1}

Let $\cR \subset \IR^2$ be a rectangle and
$\cB_1,\ldots,\cB_r\subset\,{\rm int}\, \cR$ open strictly convex
subdomains with smooth or piecewise smooth boundaries, as
described in Theorem~1.

The billiard system in $Q = \cR \setminus \cup_i \cB_i$ is
semidispersing. The collision space can be naturally divided into
two parts: \emph{dispersing}
$$
   \cM_+ = \{ (r,\varphi)\colon\, r\in \cup_{i}\, \partial\cB_i \}
$$
and \emph{neutral}
$$
   \cM_0 = \{ (r,\varphi)\colon\, r\in\partial\cR \}
$$
We set $M = \cM_+$ and consider the return map $F\colon M\to M$.
Thus we skip all the collisions with the flat boundary
$\partial\cR$ when constructing $F$.

\begin{figure}[h]
    \centering
    \psfrag{R}{$\cR$}
    \psfrag{Q}{$Q_1$}
     \includegraphics[height=2in]{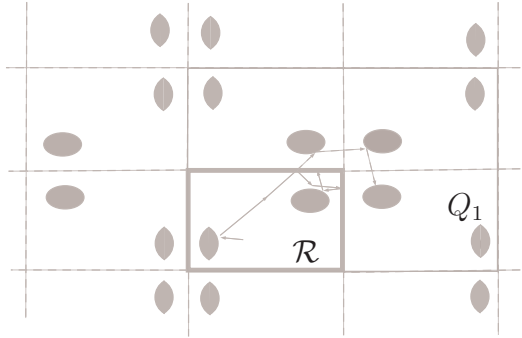}
    \renewcommand{\figurename}{Fig.}
    \caption{Unfolding of a billiard trajectory.}
    \label{FigUnfold}
\end{figure}

The map $F$ can be reduced to a (proper) collision map
corresponding to another billiard table by a standard
``unfolding'' procedure. Instead of reflecting a billiard
trajectory at $\partial\cR$ we reflect the rectangle $\cR$ with
all the scatterers $\cB_i$ across the side which our billiard
trajectory hits. Then the mirror image of the billiard trajectory
will pass straight to the new copy of $\cR$ (as if the boundary
$\partial\cR$ were transparent), see Fig.~\ref{FigUnfold}.

The mirror copies of $\cR$ obtained by successive reflections
across their sides will cover the entire plane $\IR^2$. The new
(unfolded) billiard trajectory will only hit the scatterers
$\cB_i$ and their images obtained by the mirror reflections. It is
clear that those images make a periodic structure in $\IR^2$,
whose fundamental domain $\cR_1$ consists of four adjacent copies
of $\cR$, see Fig.~\ref{FigUnfold}. In fact, $\cR_1$ can be viewed
as a 2-dimensional torus, with opposite boundaries identified.
Thus we obtain a billiards in a table $Q_1 \subset \cR_1$ with
some internal convex scatterers and periodic boundary conditions.
Our map $F$ reduces to the proper collision map in this new
billiard table.

The billiard table $Q_1$ is dispersing since all the scatterers
are strictly convex.  It has ``infinite horizon'' (or ``unbounded
horizon''), because the free paths between successive collisions
may be arbitrarily long. We call $x \in M$ an \emph{infinite
horizon point} (or, for brevity, IH-point) if its trajectory forms
a closed geodesic on the torus $\cR_1$ that only touches (but
never crosses) the boundary $\partial Q_1$, and at all points of
contact of this geodesic with $\partial Q_1$ the corresponding
components of $\partial Q_1$ lie on one (and the same) side of
this geodesic. There are at most finitely many IH-points $x \in
M$, and we denote them by $x_q$, $q \geq 1$, they play a crucial
role in our analysis.

First, let us assume that the scatterers $\cB_i$ have smooth
boundaries. Then the billiard in $Q_1$ is a classical Lorentz gas,
also known as Sinai billiard table. The exponential decay of
correlations (in fact, both statements (a) and (b) of
Theorem~\ref{tmmain}) for this billiard are proved in \cite{C99},
Section~8. But the argument in \cite{C99} is rather
model-specific. Below we outline another argument based on our
general scheme described in the previous sections. This outline
will also help us treat piece-wise smooth scatterers $\cB_i$
later. We mention some standard properties of dispersing billiards
and refer the reader to \cite{BSC90,BSC91,C97,C99} for a detailed
account.

The discontinuity curves of the map $F$ (the ``old''
singularities, in the terminology of Section~4.4) are of two
types. There are finitely many long curves dividing $M$ into
finitely many large domains. In addition, there are infinite
sequences of short singularity curves that converge to the
IH-points $x_q = (r_q, \varphi_q) \in M$ (note that $\varphi_q =
\pm \pi/2$). The singularity curves divide the neighborhoods of
the IH-points into countably many small regions (commonly called
\emph{cells}).

\begin{figure}[h]
\centering \psfrag{4}[lb][lb]{$\frac{c_2}{n}$}
\psfrag{S0}{\scriptsize$\varphi = \pi/2$} \psfrag{
5}{$\frac{c_1}{\sqrt{n}}$} \psfrag{Mn}{\scriptsize$M_{q,n}$}
\psfrag{c}{\scriptsize$x_q$} \psfrag{x}{\scriptsize$S_{q,n-1}$}
\psfrag{y}{\scriptsize$S_{q, n}$} \psfrag{a}{$(a)$}
\psfrag{b}{$(b)$}
\includegraphics[height=2in]{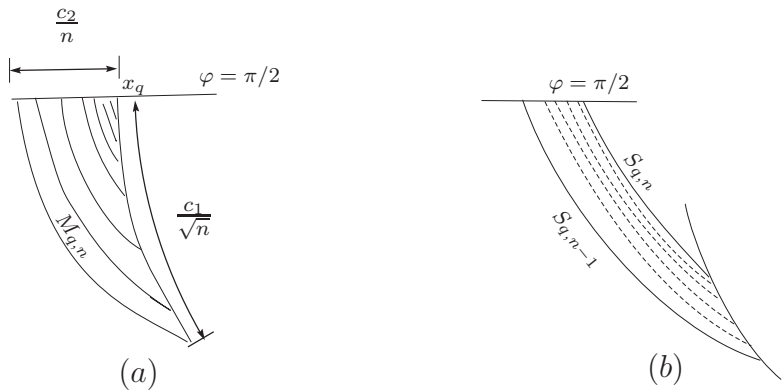}
\renewcommand{\figurename}{Fig.}
\caption{Singularity curves and cells near an
IH-point.}\label{cell1}
\end{figure}

The structure of singularity curves and cells in the vicinity of
every IH-point $x_q$ is standard -- it is shown on  the
Fig.~\ref{cell1} (a). There is one long curve $S_q$ running from
$x_q$ into $M$ and infinitely many almost parallel short curves
$S_{q,n}$, $n\geq 1$, running between $S_q$ and the border
$\partial M$ and converging to $x_q$ as $n$ goes to $ \infty$. The
dimensions of the components of this structure are indicated on
Fig. \ref{cell1} (a). We denote by $M_{q,n}$ the domain bounded by
the curves $S_{q,n}$, $S_{q,n+1}$, $S_q$ and $\partial M$ (this
domain is often called $n$-cell).

We declare all the short singularity curves $S_{q,n}$ with $n>
n_1$ secondary, where $n_1$ is sufficiently large and will be
specified below. All the other old singularity curves are declared
primary.

The ``new'' singularities (the preimages of the boundaries of the
homogeneity strips $H_k$, see Section~4.4) consist of infinite
sequences of curves that converge to the old singularities. In
particular, every curve  $S_{q,n}$ is a limit curve for an
infinite sequence of ``new'' singularity curves, see
Fig.~\ref{cell1} (b). All the new singularity curves are declared
secondary.

It is known that the complexity $K_m$ of the primary singularities
of the map $F^m$ grows at most linearly, i.e.\ $K_m \leq C_1 +
C_2m$, where $C_1,C_2 >0$ are constants (see \cite{BSC90},
Section~8). This means that no more than $C_1 + C_2m$ singularity
curves of $\cS_m$ meet at any one point $x \in M$. This implies
(\ref{Primary}) for any $\theta_1 <1$ and sufficiently large $m$.

It remains to verify the main assumption (\ref{SS}). Suppose first
that a short unstable manifold $W$ crosses new singularity curves
converging to a primary old curve. Then $W$ is divided into
countably many pieces $W_k = W \cap F^{-1}(H_k)$. The expansion of
$W_k$ under $F$ is bounded below by $\Lambda_k \geq Ck^2$, where
$C>0$ is a constant, see \cite{C99}, Equation (7.3). Thus
$$
      \sum_{k_0}^{\infty} \Lambda_k^{-1}
      \leq \sum_{k_0}^{\infty} (Ck^2)^{-1}
      \leq 2\,C^{-1}k_0^{-1},
$$
which can be made $<1/2$ by choosing $k_0$ large enough, say $k_0
> 4\, C^{-1}$.

Next, let a short unstable manifold $W$ intersect some (or all)
secondary old curves $S_{q,n}$ with some $q \geq 1$ and $n_1 \leq
n < \infty$, and near each of them it intersects infinitely many
new singularity lines converging to the old one. For each $n \geq
n_1$ denote by $W_n$ the piece of $W$ between $S_{q,n}$ and
$S_{q,n+1}$, and let $W_{n,k} = W_n \cap F^{-1}(H_k)$ for $n \geq
n_1$ and $k \geq k_0$ denote the connected components of $W
\setminus \cS$. It is important to observe that the image $F(W_n)$
only intersects homogeneity strips $H_k$ with $k \geq \chi
n^{1/4}$, where $\chi >0$ is a constant (see \cite{C99}, page
544). The expansion of $W_{n,k}$ under $F$ is bounded below by
$\Lambda_{n,k} \geq Cnk^2$, where $C>0$ is a constant, see
\cite{C99}, Equation (8.2). Thus we have
\begin{align*}
      \sum_{n=n_1}^{\infty}\
      \sum_{k=\chi n^{1/4}}^{\infty} \Lambda_{n,k}^{-1} &\leq
      \sum_{n=n_1}^{\infty}\
      \sum_{k=\chi n^{1/4}}^{\infty} (Cnk^2)^{-1} \\
      &\leq
      \sum_{n=n_1}^{\infty}\
      2\,C^{-1}\chi^{-1} n^{-5/4} \\
      &\leq
      10\,C^{-1}\chi^{-1} n_1^{-1/4},
\end{align*}
which can be made $<1/2$ by choosing $n_1$ large enough, say $n_1
> [20\, C^{-1}\chi^{-1} ]^4$.

This concludes the verification of (\ref{SS}). By
Theorem~\ref{Tm11} and the remark after Theorem~\ref{tmmain}, the
assumptions (\ref{growth1})--(\ref{growth3}) hold, hence the
induced map $F\colon M\to M$ has exponential mixing rates by
Theorem~\ref{tmmain}.

Lastly, in order to determine the rates of mixing for the original
collision map $\cF\colon \cM \to \cM$ of the billiard in $Q = \cR
\setminus \cup_i \cB_i$, we need to estimate the return times
defined by (\ref{Mpol2}). If the short singularity curves
$S_{q,n}$ are labelled in their natural order (so that $n$
increases as they approach the limit IH-point $x_q$), then
\beq \label{nMM}
   \{ x\in M\colon R(x; \cF, M) > n\} \subset
   \cup_q \cup_{i\geq cn} M_{q,i}
\eeq
where $c>0$ is a constant and $M_{q,i}$ denotes the $i$-cell. This
is a simple geometric fact, we leave the verification to the
reader.

It is easy to see on Fig.~\ref{cell1} (a) that $\mu(M_{q,n}) =
\cO(n^{-3})$, because the ``width'' of $M_{q,n}$ (its
$r$-dimension) is $\cO(n^{-2})$, its ``height'' (the
$\varphi$-dimension) is $\cO(n^{-1/2})$ and the density of the
$\mu$ measure in the $(r, \varphi)$ coordinates is $\cos\varphi =
\cO(n^{-1/2})$. Hence $\mu\bigl(R(x;\cF,M) > n\bigr) =
\cO(n^{-2})$. Theorem~\ref{tmpolycorr} now implies the required
bound on correlations
$$
   |\cC_n(f,g,\cF,\mu)|\leq\,{\rm const}\cdot (\ln n)^2/n.
$$

\begin{figure}[h]
\centering \psfrag{2}{$ \frac{c_1}{n}$}
\psfrag{3}[c][c]{\scriptsize $M_{n,q}$}
\psfrag{4}[lb][lb]{$\frac{c_2}{n}$}
 \psfrag{b1}{$(b)$}
\psfrag{a}{$(a)$} \psfrag{p}{$x_q$}
 \psfrag{yq}[rt][rt]{$x_q$}
\psfrag{b}{$b$} \psfrag{d}{$d$}
\includegraphics[height=1.5in]{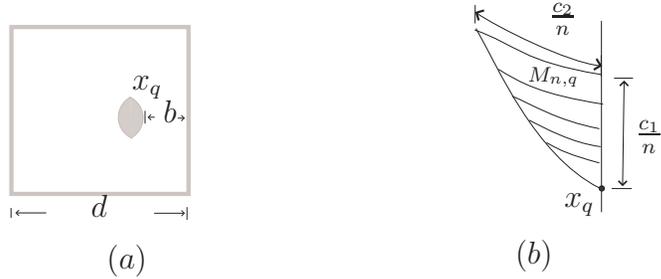}
\renewcommand{\figurename}{Fig.}
\caption{Discontinuity curves and cells near an IH2-point
$x_q$.}\label{corner}
\end{figure}

To complete the proof of Theorem~1, we need to consider scatterers
$\cB_i$ with piece-wise smooth boundaries. Now there are two types
of IH-points: those whose trajectories intersect (graze) $\partial
Q_1$ at points where $\partial Q_1$ is smooth we call them
IH1-points, and those whose trajectories intersects $\partial Q_1$
at corner points, as shown on Fig.~\ref{corner}, we call them
IH2-points.

The analysis of IH1-points is identical to the one above. The
structure of singularity curves and cells in the neighborhood of
an IH2-point $x_q = (r_q, \varphi_q)$ is shown on Fig.
\ref{corner} (b), we refer the reader to \cite{BSC90}, Section~4,
for more detail. It is important to note that $|\varphi_q| <
\pi/2$ and the first few $F$-images of a small neighborhood of
$x_p$ do not intersect $\partial M$ (they ``wander'' inside $M$,
as it is explained in \cite{BSC90}), so that there is no ``new''
singularity curves inside the cells $M_{q,n}$ for large enough
$n$.

Following \cite{BSC91}, we assume that the condition
(\ref{Primary}) holds for some $m \geq 1$ and proceed to verify
the main assumption (\ref{SS}). Let $\cK_{\min}>0$ denote the
minimum curvature of the boundary of the scatters. Let $d$ denote
the width (the smaller dimension) of $\cR$ and let
$$
   b=\min_{i\neq j} \bigl(
   {\rm dist}(\cB_i, \cB_j),\,{\rm dist}(\cB_i, \partial\cR)\bigr).
$$
Let $W$ be a short unstable manifold and denote by $W_n$ the piece
of $W$ between $S_{q, n}$ and $S_{q, n+1}$. Unstable manifolds are
represented by increasing curves in the $r\varphi$ coordinates. It
is easy to see on Fig.~\ref{corner} (b) that unstable manifolds
cannot cross infinitely many singularity curves. In fact, by a
simple geometric calculation, there exists a constant
$$
   1\leq \alpha \leq
   1+\frac{k_{\min}+\frac{\cos\varphi_q}{2b}}
   {k_{\min}+\frac{\cos\varphi_q}{d}}
$$
such that if $W$ crosses the curves $S_{q, n}$ for all $n \in
[n_1, n_2]$, then $n_2 \leq \alpha n_1$. The expansion of $W_n$
under $F$ is bounded from below by
$$
   \Lambda_n \geq 1+C n,
$$
where $C=\cK_{\min}d/\cos\varphi_q$,  see \cite{C99},
Equation~(6.8). Here we used the fact that the intercollision time
for points $x \in W_n$ is at least $d\, n$. Thus we have
\begin{align} \label{CCC}
     \sum_{m=n}^{\alpha n}
     \frac{1}{\Lambda_m} &\leq
     \frac{1}{C}\sum_{m=n}^{\alpha n}\frac{1}{m}\nonumber\\
     &\leq \frac{1}{C}\, \ln \Bigl(1+\frac{C+\eta}{1+C}\Bigr)\nonumber\\
     &\leq \frac{C+\eta}{C+C^2}
\end{align}
where $\eta=d/2b$ and we used the fact that $\ln(1+x)<x$ for
$x>0$.

The bound (\ref{CCC}) is finite, but we need it to be less than
one, which only happens if $\eta < C^2$, that is if $\cos^2\varphi_q
\leq 2bd\cK_{\min}^2$. While this is indeed the case for some
billiard tables, it is not hard to enforce the condition
\eqref{SS} for all relevant billiard tables by considering a
higher iteration of the map $F$, as it is explained in
\cite{BSC91}, page 104.

By Theorem~\ref{Tm11} and the remark after Theorem~\ref{tmmain},
the assumptions (\ref{growth1})--(\ref{growth3}) hold, hence the
induced map $F\colon M\to M$ has exponential mixing rates.

Lastly, in order to determine the rates of mixing for the original
collision map $\cF\colon \cM \to \cM$ of the billiard in $Q = \cR
\setminus \cup_i \cB_i$, we need to estimate the return times
defined by (\ref{Mpol2}). Similarly to (\ref{nMM}), we now have
$$
   \{ x\in M\colon R(x; \cF, M) > n\} \subset
   \cup_q \cup_{m\geq c_1n} M_{q,m}
$$
where $c_1>0$ is a constant.

It is easy to see on Fig.~\ref{corner} (b) that $\mu(M_{q,n}) =
\cO(n^{-3})$ for IH2-points, hence
$$
    \mu\bigl(R(x;\cF,M) > n\bigr) = \cO(n^{-2}).
$$
Theorem~\ref{tmpolycorr} now implies the required bound on
correlations
$$
   \qquad\qquad\qquad\qquad|\cC_n(f,g,\cF,\mu)|\leq\,{\rm const}\cdot (\ln n)^2/n.
   \qquad\qquad\qquad\qquad\qed
$$

\section{Proof of Theorem 2} \label{SecTm2}

Let $Q\subset \mathbb{R}^2$ be a billiard table satisfying the
assumptions of Theorem~2. Its boundary can be decomposed as
$\partial Q=\partial^+ Q \cup \partial^- Q$ so that each smooth
component $\Gamma_i\subset \partial^+ Q$ is dispersing and each
component $\Gamma_j\subset \partial^- Q$ is focusing (an arc of a
circle that is contained in $Q$). The collision space can be
naturally divided into two parts:
$$
    \cM_{\pm}=\{(r, \varphi)\,:\, r\in \partial ^{\pm}Q\}.
$$

As distinct from dispersing billiards, expansion and contraction
in $\cM_-$ are not uniform. More precisely, the expansion and
contraction are weak (per collision) during long sequences of
successive reflections in the same focussing boundary component.

These sequences become a ``disturbing factor" similar to
reflections in neutral (flat) boundary components studied in the
proof of Theorem~1, hence the induced map $F\colon M \to M$ must
be defined so that those sequences are skipped. We set
\beq \label{M2}
   M=\cM_+ \cup \{x \in \cM_-\colon\,
   \pi(x)\in \Gamma_i,\,
   \pi(\cF^{-1}x)\in \Gamma_j,
   \,j\neq i\},
\eeq
where $\pi(x) = r$ denotes the first coordinate of the point $x =
(r, \varphi)$. Observe that $M$ includes all collisions at
dispersing boundary components but only the \emph{first} collision
at every focusing component. We note that in \cite{BSC91,M04} the
\emph{last} collision (rather than the first) in every arc is used
in the construction of an induced map, but that definition of $M$
leads to unpleasant complications in the analysis, which we will
avoid here.

There are two ways in which billiard trajectories can experience
arbitrarily many reflections inside one focusing component (arc)
$\Gamma_i \subset \partial^- Q$. First, if the collisions are
nearly grazing ($\varphi \approx \pm \pi/2$), the points $x$,
$\cF(x)$, $\cF^2(x), \ldots$ are close to each other, so that the
sequence $\{ \cF^n (x) \}$ moves slowly along the arc $\Gamma_i$
until it comes to an end of the arc and escapes. This is possible
for arcs of any size.

Second, if $\varphi \approx 0$, then the billiard trajectory runs
near a diameter of $\Gamma_i$, hits $\Gamma_i$ on the opposite
side and then comes back, so that the points $x$, $\cF^2(x)$,
$\cF^4(x), \ldots$ are close to each other. Then the two sequences
$\{ \cF^{2n} (x) \}$ and $\{ \cF^{2n+1} (x) \}$ move slowly along
the arc $\Gamma_i$ until one of them finds an opening in
$\Gamma_i$ and escapes. Similarly, the trajectory can run close to
a periodic trajectory inside $\Gamma_i$ of any period $p \geq 2$.
All these, however, require the arc $\Gamma_i$ to be larger than
half-circle, which is specifically excluded by assumption (ii) of
Theorem~2. So we do not need to deal with these trajectories now,
but we will encounter them later.

\begin{figure}[h]
\centering \psfrag{n1}{\scriptsize$c_1/n$}
 \psfrag{n}{\scriptsize$c_2/n$}
\psfrag{zq}{\scriptsize$z_q$} \psfrag{p}{$\varphi$}
\psfrag{r}{$r$}
\includegraphics[height=1.5in]{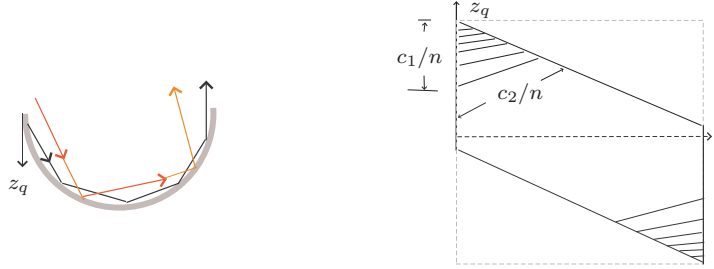}
\renewcommand{\figurename}{Fig.}
\caption{Sliding trajectories and cell structure of Bunimovich
billiard tables} \label{flowercell}
\end{figure}

In the coordinates $(r, \varphi)$, the set $M$ is the union of
rectangles and cylinders corresponding to the dispersing boundary
components and parallelograms corresponding to the focusing
boundary components, as the one shown on Fig.~\ref{flowercell}. In
each parallelogram, unstable manifolds are represented by
decreasing curves in the $(r,\varphi)$ coordinates, and
singularity manifolds -- by increasing curves (which is opposite
to dispersing billiards). Between successive collisions at the
focusing boundary, each unstable manifold first converges
(focuses, or collapses), passes through a conjugate (defocusing)
point, and then diverges. We refer the reader to
\cite{Bu74,Bu79,CM} for a detailed account of hyperbolicity in
Bunimovich billiards. It is important that the conjugate point
always lies in the first half of the segment between the
consecutive reflections, which guarantees a monotonic expansion of
unstable manifolds from collision to collision in a special
$p$-metric on unstable curves, defined by $dp = \cos \varphi \,
dr$, see \cite{BSC90,BSC91} for definitions and details.

The induced map $F$ has infinitely many discontinuity lines, which
accumulate in the neighborhood of two vertices (top and bottom) of
each parallelogram, see Fig.~\ref{flowercell}. Each such vertex
$z_q$ is a limit point for infinitely many almost parallel
straight segment $S_{q, n} \subset \cS$ running between two
adjacent sides of the parallelogram. They divide the neighborhood
of $z_q$ into countably many cells, which we denote, as before, by
$M_{q, n}$, $n \geq 1$. Here the cell $M_{q,n}$ consists of points
experiencing exactly $n$ consecutive collisions with the arc
(counting the first collision included in $M$). In fact, for $n$
large enough, those trajectories are more like ``sliding'' along
the arc rather than reflecting off it, hence the corresponding
singular points are said to be of \emph{sliding type}, see more
details in \cite{BSC90} and \cite{BSC91}.

We declare all the short singularity curves $S_{q,n}$ with $n>
n_1$ secondary, where $n_1$ is sufficiently large and will be
specified below. All the other singularity curves are declared
primary.

Following \cite{BSC91}, we assume that the condition
(\ref{Primary}) holds for some $m \geq 1$. We remark that
(\ref{Primary}) holds for generic billiard tables, i.e.\ for an
open dense set of billiard tables in the $C^3$ metric, which one
can show by using standard perturbation techniques, but we do not
pursue this goal here.

We now proceed to verify the main assumption (\ref{SS}). Let $W
\subset M_{q,n}$ be an unstable manifold, so that $F=\cF^n$ on
$W$. Then the map $\cF^{n-1}$ expands $W$ by a factor $\cO(n)$,
and then $\cF$ expands the manifold $\cF^{n-1}(W)$ by a factor
$\cO(n)$, see \cite{BSC91}, Section~2.5. Thus, the induced map $F$
expands $W$ by a factor $\geq cn^2$, where $c>0$ is a constant,
and we have
$$
   \sum_{m=n_1}^{\infty}\frac{1}{\Lambda_m}
   \leq \sum_{m=n_1}^{\infty}\frac{1}{cm^2}
   \leq \frac{2}{cn_1},
$$
which can be made $<1/2$ by choosing $n_1$ large enough, say $n_1
> 4/c$.

By Theorem~\ref{Tm11} and the remark after Theorem~\ref{tmmain},
the assumptions (\ref{growth1})--(\ref{growth3}) hold, hence the
induced map $F\colon M\to M$ has exponential mixing rates.

Lastly, we need to estimate the return times (3.6) to determine
the rates of mixing for the original collision map $\cF \colon \cM
\to \cM$. The cell $M_{q,n}$ has ``width'' $\cO(n^{-1})$,
``height'' $\cO(n^{-2})$, see Fig.~\ref{flowercell}, and the
density of the measure $\mu$ on $M_{q,n}$ is $\cO(n^{-1})$, hence
$\mu(M_{q,n})=\cO(n^{-4}).$ Since
$$
    \{x\in M\colon\, R(x;\cF, M)>n\}\subset
    \cup_q \cup_{m>n} M_{q,m},
$$
we get $\mu\bigl(R(x;\cF, \cM)>n\bigr)=\cO(n^{-3})$. Theorem 4 now
implies the  required bound on correlations
$$
   |\cC_n(f,g,\cF, \mu)|\leq \,{\rm const}\cdot(\ln n)^3/n^2.
$$
Note that the faster rate of the decay of correlations here (as
compared to Theorem~1) is purely due to the smaller cells
$M_{q,n}$, and not due to any properties of the induced map $F$.
\qed

  \section{Proof of Theorem 3}

Let $Q\subset \mathbb{R}^2$ be a stadium. Its boundary can be
decomposed as
$$
   \partial Q=\partial^0 Q \cup \partial^- Q,
$$
where $\partial Q^0=\Gamma_1\cup \Gamma_2$ is the union of two straight sides of $Q$,
and $\partial Q^-=\Gamma_3\cup \Gamma_4$ is the union of two arcs. The collision space
can be naturally divided into focusing and neutral parts:
$$
   \cM_{0}=\{(r, \varphi)\,:\, r\in \partial^0 Q\},
   \quad\quad
   \cM_{-}=\{(r, \varphi)\,:\, r\in \partial^- Q\}.
$$
We define the induced map $F\colon M\to M$ on the set
\beq \label{M3}
   M= \{x\in \cM_-\colon\, \pi(x)\in \Gamma_i,\,
   \pi(\cF^{-1}x)\in \Gamma_j, \,j\neq i\},
\eeq
Note that $M$ only contains the \emph{first} collisions with the
arcs (the collisions with the straight lines are skipped
altogether). In the coordinates $(r, \varphi)$ the set $M$ is the
union of two parallelograms.

First let us consider the straight stadium, see  Fig.~\ref{semi}
(c). This model was already handled by Markarian \cite{M04}, but
we present here a simplified proof based on our general method.

\begin{figure}[h]
\centering \psfrag{fx}{$F(z)$} \psfrag{x}{$x$} \psfrag{y}{$y$}
\psfrag{x1}{ $x_1$} \psfrag{y1}{$y_1$} \psfrag{z}{$z$}
\includegraphics[height=1.5in]{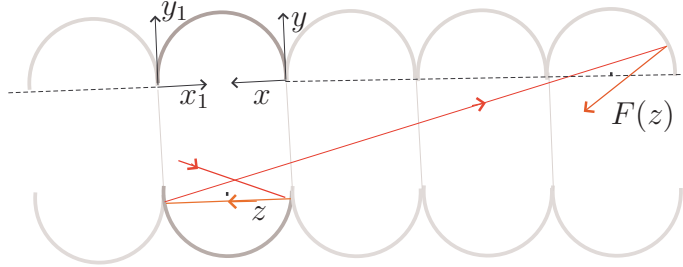}
\renewcommand{\figurename}{Fig.}
\caption{Unfolding of straight stadium}\label{unstadium}
\end{figure}

The map $F$ can be viewed as a collision map corresponding to
another billiard table obtained by ``unfolding" the stadium along
the straight boundary segments. In other words, instead of
reflecting a billiard trajectory at $\partial^0 Q$, we reflect the
stadium across its straight boundaries, see Fig.~\ref{unstadium}.

Then the mirror image of the billiard trajectory will pass
straight through $\partial Q$ until it meets the mirror image of
the arcs and get reflected there. The mirror copies of the
straight stadium obtained by successive reflections about their
straight sides make an unbounded strip, see Fig.~\ref{unstadium}.
The new billiard in the unbounded strip has ``infinite horizon'',
since the free path between collisions may be arbitrarily long.

\begin{figure}[h]
\centering \psfrag{A}[rb][rb]{$y_1$} \psfrag{B}{$y_2$}
\psfrag{C}{$x_1$} \psfrag{D}{$x_2$} \psfrag{n}{\scriptsize$c/n$}
\psfrag{c1}{\scriptsize$c_1/n$} \psfrag{(a)}{$(a)$}
\psfrag{(b)}{$(b)$} \psfrag{1}{\scriptsize$M_{n}^2$}
\psfrag{2}[c][c]{\scriptsize$M_{n}^1$} \psfrag{r}{$r$}
\includegraphics[height=2in ]{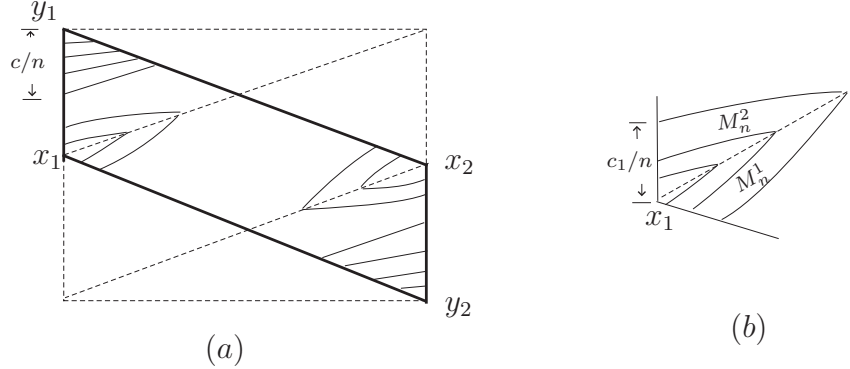}
\renewcommand{\figurename}{Fig.}
\caption{Discontinuity curves  of straight stadium }\label{stadiumcell}
\end{figure}

The singularity set $\cS \subset M$ of the map $F$ consists of two
types of infinite sequences of singularity curves, as shown on
Fig.~\ref{stadiumcell}. The first type accumulate near the top and
bottom vertices of the parallelograms, they are generated by
trajectories nearly ``sliding'' along the circular arcs. These are
identical to the ones discuss in the proof of Theorem~2, so we
omit them.

The curves of the second type accumulate near the other two
vertices of the parallelograms (those lie on the line
$\varphi=0$), they are generated by trajectories experiencing
arbitrary many bounces between the two straight sides of the
stadium (i.e.\ by long collision-free flights in the unbounded
strip). The structure of those singularity curves is shown on
Fig.~\ref{stadiumcell} (b), see \cite{BSC90}, Section~6.3, for
more details.

Actually, there are two types of cells near the vertices $x_1$ and
$x_2$ in $M$. The first type contains points mapped (by $\cF$)
directly to a straight side of the stadium, we denote them by
$M_n^1$, $n \geq 1$; the second type contains points that are
mapped by $\cF$ onto the same arc first and then to a straight
side, we denote them by $M_n^2$, $n \geq 1$. In
Fig.~\ref{unstadium}, the point $z$ belongs to $M_4^2$ and $F(z)$
belongs to $M_2^1$. Points in $M_n^1$ and $M_n^2$ experience
exactly $n$ reflections off the straight sides before landing on
the opposite arc of $\partial Q$.

To prove the condition (\ref{Primary}) it is enough to establish
the linear growth of the complexity
\beq \label{linK}
   K_{{\rm P},m} \leq C_1+C_2m,
\eeq
which is similar to the one obtained for the Lorentz gas in
\cite{BSC90}, Section~8 (recall our proof of Theorem~1). The proof
of (\ref{linK}) for the Lorentz gas is based on the continuity of
the billiard flow (as it is explained in \cite{C01}, Lemma~5.2).
The flow in the stadium is obviously continuous as well, thus the
same argument implies the linear estimate (\ref{linK}) for the
stadium, we omit details.

We now proceed to verify the main assumption (\ref{SS}). Our
calculations are based on two known facts mentioned in
\cite{BSC91}, Section 2.6, which can be verified by direct
calculations. First, if an unstable manifold $W$ intersects cells
$M_n^1$ (and $M_n^2$) with $n_1 < n < n_2$, then
$$
    n_2 \leq 9n_1 + \,{\rm const}.
$$
Second, the expansion factor of $F$ on each piece $W\cap M_n^1$,
$n \geq 1$, is $\gtrsim 4n$ and the expansion factor on each piece
$W\cap M_n^2$ is $\gtrsim 8n$. Thus we have
$$
   \sum_{m=n_1}^{n_2}\frac{1}{\Lambda_m}\leq
   \sum_{m=n_1}^{n_2}\Bigl(\frac{1}{4m}+\frac{1}{8m}\Bigr)
   \leq \frac 38\, \ln 9 < 1.
$$
By Theorem~\ref{Tm11} and the remark after Theorem~\ref{tmmain},
the assumptions (\ref{growth1})--(\ref{growth3}) hold, hence the
induced map $F\colon M\to M$ has exponential mixing rates.

Next we need to estimate the return times (3.6) to determine the
rates of mixing for the original collision map $\cF \colon \cM \to
\cM$. The $n$-cells near the top and bottom vertices $y_1$ and
$y_2$ have measure $\cO(n^{-4})$, see the previous section. The
$n$-cells $M_n^1$ and $M_n^2$ have measure $\cO(n^{-3})$, as one
can easily see from Fig.~\ref{stadiumcell} (b) (note that $\varphi
\approx 0$, hence $\cos\varphi \approx 1$). Thus we get
$\mu\bigl(R(x;\cF, \cM)>n\bigr)=\cO(n^{-2})$. Theorem 4 now
implies the  required bound on correlations
$$
   |\cC_n(f,g,\cF, \mu)|\leq \,{\rm const}\cdot(\ln n)^2/n.
$$
Note that there are two types of cells here, and the rate of the
decay of correlations is determined by the ``worse'' (larger)
cells $M_n^1$ and $M_n^2$ (those near the points $x_1$ and $x_2$).

This concludes the proof of Theorem~3 for the straight stadium.
\qed \medskip

\begin{figure}[h]
\centering \psfrag{x1}{$x_1$} \psfrag{y1}{$y_{1}$}
\psfrag{x}{$x_2$} \psfrag{y}{$y_2$} \psfrag{z}{\scriptsize{$Fz$}}
\psfrag{fz}{\scriptsize{$F^2z$}} \psfrag{z1}{\scriptsize$\cF^n z$}
\psfrag{a}{$(a)$} \psfrag{b}{$(b)$}
\includegraphics[height=2.5in ]{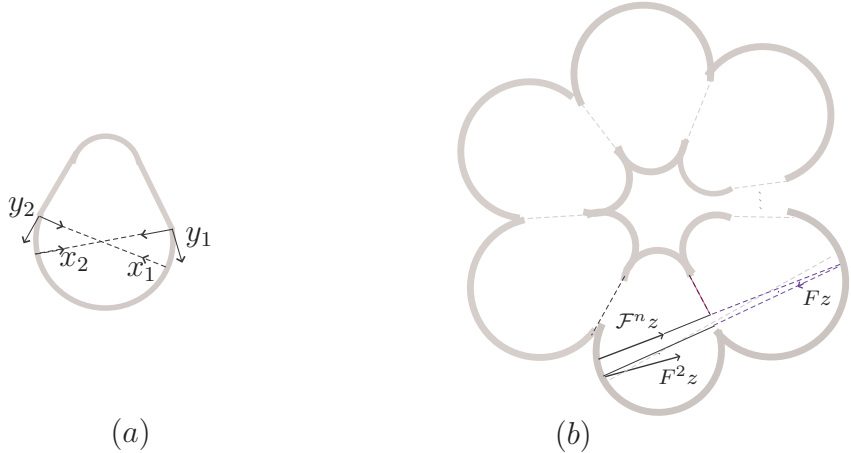}
\renewcommand{\figurename}{Fig.}
\caption{Unfolding of a skewed stadium}\label{unstadium2}
\end{figure}

Next we consider a skewed stadium (or a ``drive-belt'' table), see
Fig.~\ref{unstadium2} (a). Again, we can unfold the skewed stadium
by reflecting it repeatedly along the flat boundaries, as shown on
Fig.~\ref{unstadium2} (b). The new billiard table has similar
structure to the Bunimovich type billiard tables of the previous
section, but it does not satisfies assumption (ii) of Theorem 2,
since it necessarily contains an arc larger than half a circle.

The ``unfolding" of skewed stadium  also shows that it has
``finite horizon'', since the free path between successive
collisions is uniformly bounded from above, thus there are no
cells like $M_n^1$ and $M_n^2$ described above, but there is a
new, equally influential, type of cells, see below.

In the coordinates $(r, \varphi)$ the set $M$ is the union of two
parallelograms, one corresponds to the smaller arc, the other to
the larger arc, we call them \emph{small} and \emph{big}
parallelograms, respectively.

The singularity set $\cS \subset M$ of the map $F$ consists of two
types of infinite sequences of singularity curves. Curves of the
first type accumulate near the top and bottom vertices of each
parallelogram, they are generated by trajectories nearly
``sliding'' along the circular arcs. These are identical to the
ones discuss in the proof of Theorem~2, so we omit them again.

\begin{figure}[h] \centering \psfrag{1}
{\scriptsize$z$} \psfrag{2}{\scriptsize$y$}
\psfrag{1'}{\scriptsize$Fz$} \psfrag{2'}{\scriptsize$Fy$}
\psfrag{1''}{{\scriptsize$F^2z$}} \psfrag{M}{{\footnotesize$M$}}
\psfrag{x}{\footnotesize$x_1$}
\psfrag{y}{\footnotesize$x_2$}
\psfrag{a}{\footnotesize$S_1$}
\psfrag{b}{\footnotesize$S_2$} \psfrag{F}{\footnotesize$F$}
\psfrag{c}{\scriptsize$c/n$}
\includegraphics[height=2in]{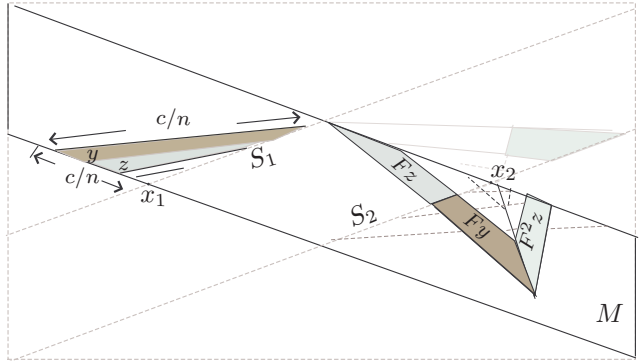}
\renewcommand{\figurename}{Fig.}
\caption{ The  cell structure of skewed stadium
}\label{skewstadiumcell}
\end{figure}

Curves of the second type accumulate near two points $x_1$ and
$x_2$ on the slanted sides of the big parallelogram (where those
sides intersect the line $\varphi = 0$). There are two parallel
long singularity lines $S_1$ and $S_2$ starting at $x_1$ and
$x_2$, respectively, and running into $M$, and two infinite
sequences of almost parallel straight segments $S_{q,n}$, where
$q=1,2$ and $n \geq 1$, running between $S_q$ and the nearby side
of $M$ and converging to $x_q$ as $n \to \infty$, see
Fig.~\ref{skewstadiumcell}.

The singularities of this second type are generated by
trajectories experiencing arbitrary many collisions with the large
arc while running almost along its diameter (this type of
trajectories was described in the previous section, but not
studied in detail there, because it was ruled out by the
assumption (ii) of Theorem~2). Denote by $M_{q,n}$ the $n$-cell
bounded by $S_{q,n}$, $S_{q,n+1}$, $S_q$ and $\partial M$. The
$n$-cell $M_{q,n}$ consists of points experiencing exactly $n$
collisions with the large arc.

There is a peculiar feature of the new cells $M_{q,n}$. Points in
the upper half of $M_{q,n}$ near $x_q$ (the dark grey area marked
by $y$ on Fig.~\ref{skewstadiumcell}) are mapped by $F$ into the
cells $M_{3-q,m}$, $n_1 \leq m \leq n_2$, near the other limit
point $x_{3-q}$, and $n_2 \leq 49n_1 + \,{\rm const}$ (see the
dark grey area marked by $Fy$). However, points in the lower half
of the cell $M_{q,n}$ (the light grey area marked by $z$ on
Fig.~\ref{skewstadiumcell}) are mapped under $F$ into a
neighborhood of the other limit point $x_{3-q}$ above or below the
line $S_{3-q}$ without crossing any singularity lines there (see
the light grey area marked by $Fz$), and the second iterate of
$F$, as it is illustrated by Fig.~\ref{unstadium2} (b), maps them
into the cells $M_{3-q,m}$, $n_1 \leq m \leq n_2$ with $n_2 \leq
49n_1 + \,{\rm const}$ (see the light grey area marked by $F^2z$).
The bound $n_2 \leq 49n_1 + \,{\rm const}$ follows from elementary
(but tedious) calculations which we leave out.

Fig.~\ref{unstadium2} (b) shows the dynamics of the lower half of
the cell $M_{q,n}$. Any point $z$ in that part experiences $n$
collisions with the large arc, then crosses a straight side of $Q$
and lands on the adjacent copy of the large arc (this becomes
$Fz$), but then it crosses the same straight side back and lands
on the old large arc again, where it starts another long series of
$m$ reflections, $n_1 \leq m \leq n_2$, running nearly along the
arc's diameter.

The above analysis implies that if an unstable manifold $W$
intersects cells $M_{q,n}$ with $n_1 \leq n_2$, then $n_2 \leq
49n_1 + \,{\rm const}$. Next, by a direct calculation (we omit
details) the expansion factor of the map $F$ on the curve  $W \cap
M_{q,n}$ is $\gtrsim 8n$. However, since $W_n = W \cap M_{q,n}$
consists of two parts (the upper half, call it $W_n'$, and the
lower half, call it $W_n''$), which evolve differently, the number
of pieces of $W$ in our estimate is doubled and we get
$$
    \sum_{m=n_1}^{n_2} \frac{1}{\Lambda_m'}
    + \frac{1}{\Lambda_m''}
    \leq 2\sum_{m=n_1}^{n_2} \frac{1}{8m}
    \eqsim \frac{1}{4}\,\ln 49<1.
$$
This finishes the proof of (\ref{SS}). Now by Theorem~\ref{Tm11}
and the remark after Theorem~\ref{tmmain}, the assumptions
(\ref{growth1})--(\ref{growth3}) hold, hence the induced map
$F\colon M\to M$ has exponential mixing rates.

Lastly, we need to estimate the return times (3.6) to determine
the rates of mixing for the original collision map
$\cF:\cM\to\cM$. The cells near the top and bottom vertices of the
parallelograms of $M$ have measure $\cO(n^{-4})$, see the previous
section. The $n$-cells $M_n$ have measure $\cO(n^{-3})$, as one
can easily see from Fig.~\ref{skewstadiumcell} (note again that
$\varphi \approx 0$, hence $\cos\varphi \approx 1$). Thus we get
$\mu\bigl(R(x;\cF, \cM)>n\bigr)=\cO(n^{-2})$. Theorem~4 now
implies the  required bound on correlations
$$
   |\cC_n(f,g,\cF, \mu)|\leq \,{\rm const}\cdot(\ln n)^2/n.
$$
The proof of Theorem~3 is complete. \qed

\section{Other examples and open questions}

Here we discuss two types of chaotic billiard tables for which our
method fails. This discussion will demonstrate the limitations of
the method, in its present form, and indicate directions for
future work.

First, let $Q$ be a Bunimovich type table satisfying the
assumption (i), but not (ii), of Theorem~2, that is let
$\partial^- Q$ contain an arc $\Gamma_i$ larger than half a
circle. Suppose we define the induced map $T\colon M\to M$ by
(\ref{M2}) again, then $M$ will contains a parallelogram $M_i$
corresponding to the arc $\Gamma_i$, and the structure of the
singularity lines in $M_i$ will be the same as the one shown on
Fig.~\ref{skewstadiumcell}. In particular, the expansion factor of
$F$ on any unstable manifold $W \cap M_{q,n}$ will be $\sim cn$,
where $c>0$ is a constant.

However, unlike the case of a drive-belt region illustrated on
Fig.~\ref{skewstadiumcell}, now some unstable manifolds may
intersect infinitely many cells $M_{q,n}$, in particular all
$n$-cells with $n \geq n_1$ for some $n_1>0$. In that case
\beq \label{inf}
    \sum_{m=n_1}^{\infty} \frac{1}{\Lambda_m}
    \sim \sum_{m=n_1}^{\infty} \frac{1}{cm} = \infty,
\eeq
and so the condition (\ref{SS}) fails. And it fails in a major
way, since neither $F$ nor any power $F^m$ can possibly satisfy
(\ref{SS}).

This failure poses interesting questions. Does this mean that the
induced map $F\colon M\to M$ has subexponential decay of
correlations? If not, can the method be improved to overcome the
trouble and establish exponential mixing for $F$? Or should one
seek a different definition of the induced map $F$ in order to
avoid the trouble? These questions remain open at the moment.

A similar failure occurs for a special modification of the
stadium, where $Q$ is bounded by two parallel straight segments
and two arcs which are less than half a circle (in that case the
arcs have to be transversal to the segments at the intersection
points). For the hyperbolicity and ergodicity of this model, see
\cite{BSC90}, Section~6.3. Suppose we define the induced map
$F\colon M\to M$ by (\ref{M3}), as before. Then one can
investigate the singularities of this map and find (see
Section~6.3 of \cite{BSC90} for more details) that an unstable
manifold $W$ can be broken by $\cS$ into infinitely many pieces
$W_n$, $n\geq n_1$, and the expansion factor of the map $F$ on
$W_n$ is $\sim cn$, where $c>0$ is a constant. Hence again we
arrive at a divergent series similar to (\ref{inf}), so the
condition (\ref{SS}) fails. It is not clear what this trouble
implies for the map $F$ and how to deal with it.

These two example demonstrate the limitations of our algorithm, in
its present form. There are other models, not discussed in this
paper, to which our scheme probably applies but it requires
substantial extra effort. One of them is a dispersing billiard
table with cusps (corner points where the adjacent boundary
components are tangent to each other). The estimation of mixing
rates for this model remains an open problem since the first
publication \cite{Ma} on it in 1983. The other is a dispersing
billiard table where the curvature of the boundary vanishes at
some points, i.e.\ where the boundary looks like the graph of
function $y = c|x|^{\beta}$ near $x=0$ with some $\beta > 2$ (then
the curvature vanishes at $x=0$). Our preliminary calculations
show that the rate of mixing is polynomial and its degree depends
on $\beta$. The work on both models is currently underway.

Another interesting question is how to bound the correlations from
below (recall that all our bounds are from above). It is
intuitively clear from the results of \cite{Y99} that the
polynomial bounds on correlations we established here (up to the
logarithmic factor, though) are sharp, i.e.\ cannot be improved.
However, formal lower bounds on correlations may be hard to
obtain. We only point out recent results by O.~Sarig in this
direction \cite{Sa}.

\end{document}